\documentclass[preprint, a4paper, superscriptaddress, aps, pra]{revtex4-1}

\usepackage{latexsym}
\usepackage{natbib}
\usepackage{graphicx}
\usepackage{amsmath}
\usepackage{mathcomp}
\usepackage{amssymb}
\usepackage{ulem}
\usepackage{gensymb}
\usepackage{lineno} 
\usepackage{lipsum}
\usepackage{siunitx}

\usepackage{braket}

\usepackage[usenames,dvipsnames]{xcolor} % for named colors

\newcommand{\code}[1]{\texttt{#1}} % highlight code names

\begin{document}

\title{Magnetic domain walls of the van der Waals material Fe$_3$GeTe$_2$}

%\title{Spin-polarized scanning tunneling microscopy of van der Waals magnet Fe$_3$GeTe$_2$}

\author{Hung-Hsiang Yang}
\affiliation{Physikalisches Institut, Karlsruhe Institute of Technology, 76131 Karlsruhe, Germany}

\author{Namrata Bansal}
\affiliation{Physikalisches Institut, Karlsruhe Institute of Technology, 76131 Karlsruhe, Germany}

\author{Philipp R\"u{\ss}mann}
\affiliation{Peter Gr\"unberg Institut (PGI-1) and Institute for Advanced Simulation (IAS-1) Forschungszentrum J\"ulich GmbH, D-52425 J\"ulich}

\author{Markus Hoffmann}
\affiliation{Peter Gr\"unberg Institut (PGI-1) and Institute for Advanced Simulation (IAS-1) Forschungszentrum J\"ulich GmbH, D-52425 J\"ulich}

\author{Lichuan Zhang}
\affiliation{Peter Gr\"unberg Institut (PGI-1) and Institute for Advanced Simulation (IAS-1) Forschungszentrum J\"ulich GmbH, D-52425 J\"ulich}

\author{Dongwook Go}
\affiliation{Peter Gr\"unberg Institut (PGI-1) and Institute for Advanced Simulation (IAS-1) Forschungszentrum J\"ulich GmbH, D-52425 J\"ulich}

\author{Qili Li}
\affiliation{Physikalisches Institut, Karlsruhe Institute of Technology, 76131 Karlsruhe, Germany}

\author{Amir-Abbas Haghighirad}
\affiliation{Institute for Quantum Materials and Technologies, Karlsruhe Institute of Technology, 76131 Karlsruhe, Germany}

\author{Kaushik Sen}
\affiliation{Institute for Quantum Materials and Technologies, Karlsruhe Institute of Technology, 76131 Karlsruhe, Germany}

\author{Stefan Bl\"ugel}
\affiliation{Peter Gr\"unberg Institut (PGI-1) and Institute for Advanced Simulation (IAS-1) Forschungszentrum J\"ulich GmbH, D-52425 J\"ulich}

\author{Matthieu Le Tacon}
\affiliation{Institute for Quantum Materials and Technologies, Karlsruhe Institute of Technology, 76131 Karlsruhe, Germany}

\author{Yuriy Mokrousov}
\affiliation{Peter Gr\"unberg Institut (PGI-1) and Institute for Advanced Simulation (IAS-1) Forschungszentrum J\"ulich GmbH, D-52425 J\"ulich}
\affiliation{Institute of Physics, Johannes Gutenberg-University Mainz, 55099 Mainz, Germany}

\author{Wulf Wulfhekel}
\affiliation{Physikalisches Institut, Karlsruhe Institute of Technology, 76131 Karlsruhe, Germany}
\affiliation{Institute for Quantum Materials and Technologies, Karlsruhe Institute of Technology, 76131 Karlsruhe, Germany}

\date{\today}

\begin{abstract}

Among two-dimensional materials, Fe$_3$GeTe$_2$ has come to occupy a very important place owing to its ferromagnetic nature with one of the highest Curie temperatures among known van der Waals materials and the potential for hosting skyrmions. In this combined experimental and theoretical work, we investigate the magnetic bubble domains as well as the microscopic domain wall profile using spin-polarized scanning tunneling microscopy in combination with atomistic spin-dynamics simulations performed with parameters from density functional theory calculations. We find a weak magneto-electric effect influencing the domain wall width by the electric field in the tunneling junction and determine the critical magnetic field for the collapse of the bubble domains. Our findings shed light on the origins of complex magnetism that Fe$_3$GeTe$_2$ exhibits.

\end{abstract}

\maketitle

\section*{INTRODUCTION}

Two-dimensional (2D) van der Waals (vdW) materials offer an ideal playground for fundamental and applied science due to their unique electronic structures and their ability to build complex stacks by exfoliation.
Recently discovered 2D magnetism in vdW materials containing 3$d$ transition-metal atoms \cite{ma2012evidence,lebegue2013two,Tong2016} extends the range of application of vdW materials in spintronics. Depending on the metal ions and stacking, various intra- and interlayer magnetic interactions (ferro-, antiferromagnetic or frustrated) combined with a metallic or insulating band structure were predicted and experimentally found, allowing spin injection and detection, electrical manipulation of magnetism, spin-orbit driven magnetic topology and entirely new fields such as spin-valleytronics \cite{Tong2016}.
Multiple groups using different techniques verified magnetism in CrI$_3$ ($T_C$=45 K) \cite{Huang2017, Jiang2018}, Cr$_2$Ge$_2$Te$_6$ ($T_C$=30 K) \cite{Gong2017}, and Fe$_3$GeTe$_2$ (FGT) \cite{Deng2018, fei2018two, Kim2018, nguyen2018visualization, ding2019observation}. For an overview including more than 60 different magnetic vdW materials, we refer to reference \cite{shabbir2018long}. Among magnetic 2D vdW materials, FGT attracts significant attention due to its high ferromagnetic transition temperature of 150-220 K \cite{Kim2018} and its alleged semi-metallic nature \cite{nguyen2018visualization}. Additionally, recent reports demonstrated that FGT could host magnetic skyrmion bubbles \cite{nguyen2018visualization, ding2019observation}. Experimental efforts were made to extract the important magnetic characteristics of FGT such as the saturation magnetization $M_s$ and uniaxial magnetocrystalline anisotropy constant $K$ via superconducting quantum interference device magnetometry \cite{leon2016magnetic}. The other essential property is the exchange constant $A$, which is estimated from the calculated spin stiffness to be in the order of $10^{-12} J \cdot m^{-1}$ \cite{costa2020nonreciprocal}. A direct way to experimentally determine the exchange constant is to measure the magnetic domain wall width, which is proportional to  $\sqrt{|A/K|}$ \cite{hubert-schafer2000}. In the case of FGT, the domain-wall width is, however, expected to be in the order of nanometers, which makes it experimentally challenging to provide precise estimates.

In this work, we investigate the FGT surface using spin-polarized scanning tunneling microscopy (SP-STM) with antiferromagnetic tips. The spin-polarized density of states is revealed by spin-polarized scanning tunneling spectroscopy measurements and density functional theory (DFT) calculations. Using SP-STM, we measure the domain-wall width with sub-nanometer resolution and compare it to atomistic spin-dynamics simulations based on the exchange parameters extracted from our DFT calculations. 
We find that the size of bubble domains monotonically decreases with the application of a perpendicular magnetic field opposing the bubble magnetization until the bubble collapses at a critical field of about 0.32 T. With these results, we put a solid base for the understanding of the complex magnetism of FGT especially regarding for the use of the topologically non-trivial magnetic states of the material in hybrid structures. 

\section*{RESULTS and Discussion}

\begin{figure}[ht!]
 \includegraphics[width=1\linewidth]{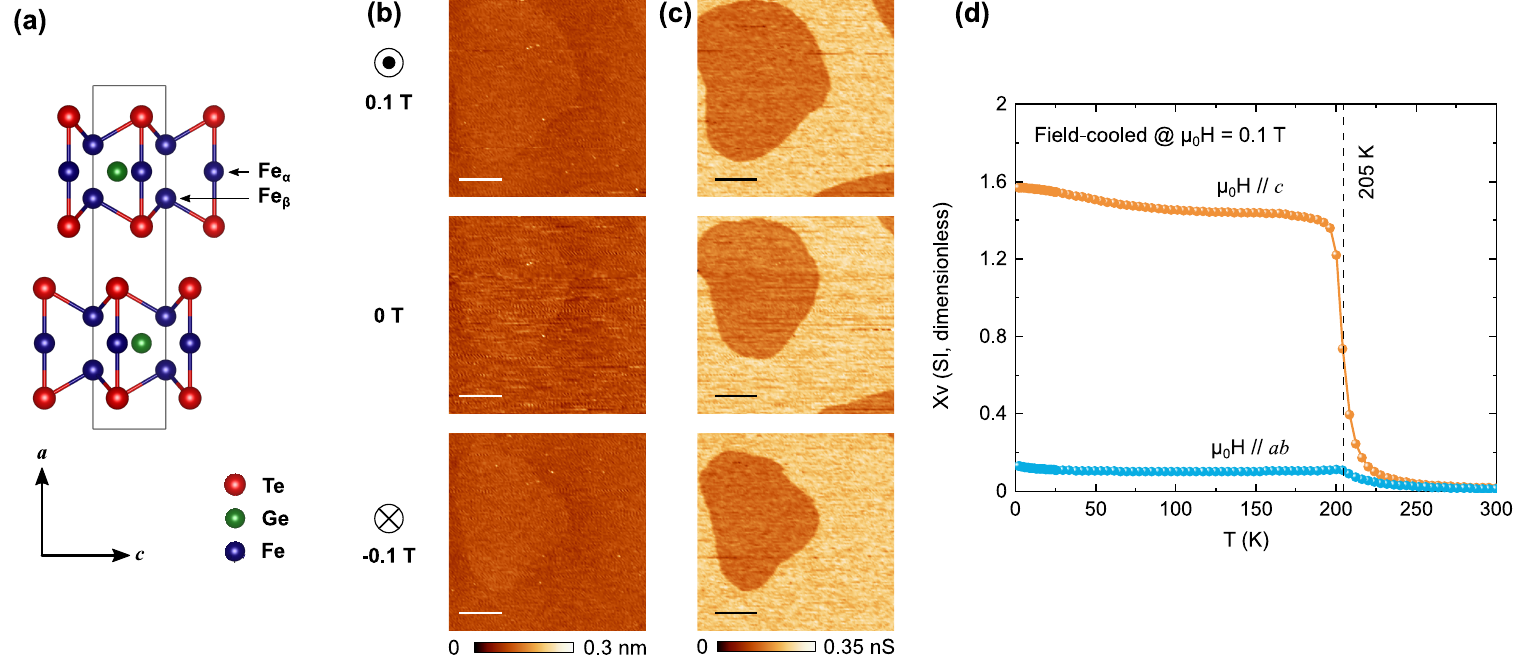}
\caption{(a) Crystal structure of Fe$_3$GeTe$_2$. (b) STM images of FGT at 0.1 T (upper), 0 T (middle), and $-$0.1 T (lower panel). (c) $dI/dU$ maps taken simultaneously with (a). (d) Magnetic volume susceptibility(${\chi}_{\nu}$) as a function temperature measured in field-cooled condition (0.1 T). Scale bar indicates 500 nm. Imaging conditions: sample bias voltage ($U$) : $-$500 mV, tunneling current ($I$) : 50 pA, and modulation voltage ($U_{mod}^{rms}$) : 80 mV.}
\label{Fig1}
\end{figure}

Fe$_3$GeTe$_2$ is a known pseudo-2D transient ferromagnetic (FM) metal \cite{Deiseroth2006, chen2013magnetic} and its unit cell is composed of two Fe$_3$Ge layers sandwiched between two Te layers which associates with P6$_3$/mmc space group ($a$ = 3.991 {\AA}, $c$ = 16.33 {\AA}). The two disparate Wyckoff positions are being occupied by two Fe atoms designated as Fe$_{\alpha}$ and Fe$_{\beta}$ (figure~\ref{Fig1}(a)). Fe$_{\alpha}$ atoms are situated in a hexagonal net in a layer with only Fe atoms. Fe$_{\beta}$ and Ge atoms are covalently bound in an adjacent layer. The Fe$_3$Ge layers are separated by van der Waals gapped Te double layers. In recent studies, magnetic force microscopy has been used to probe the magnetic domains of FGT \cite{fei2018two}. Magnetic domains of bubble-shaped and stripe shape have been detected at low temperatures \cite{leon2016magnetic}. 

In our study, FGT samples were grown by chemical vapour transport method and were transferred into UHV followed by in-situ cleavage at room temperature (see Methods for details) followed by STM investigation at $\approx$ 0.7 K. The surfaces of the bulk FGT sample (thickness of the order of 200 $\mu$m) prepared this way were extremely flat showing terraces of extensions of over 2 $\mu$m. Figure~\ref{Fig1}(d) shows magnetic volume susceptibility (${\chi}_{\nu}$) as a function of temperature ($T$) for $H \parallel c$ and $H \parallel ab$. ${\chi}_{\nu}$-vs-T clearly indicates that FGT is a ferromagnet and has a critical temperature ($T_c$) of 205 K. Furthermore, ${\chi}_{\nu}$ strongly depends on the direction of the applied magnetic field ($H$). We find that the ${\chi}_{\nu}$ is 15 times larger at $H \parallel c$ than the one at $H \parallel ab$ indicating that the magnetic easy axis is along $c$-axis, i.e.\ Fe-moments prefer to align along $c$-axis. 

First, we focus on the magnetic contrast and the spin polarization of the states above the surface.
The magnetic domains at the Fe$_2$GeTe$_3$ surface were mapped using the spin-polarized scanning tunneling microscopy (SP-STM) with a Cr-coated W tip, which has out-of-plane sensitivity at a Cr coverage lower than 50 monolayers \cite{Wachowiak2002a}. Figure~\ref{Fig1}(b) shows topographic SP-STM images of the FGT surface displaying faint domains. These are due to the tunneling magnetoresistance (TMR) effect \cite{Julliere1975} slightly changing the tunneling probability and thus affecting the $z$-position in the constant current feedback mode \cite{Wiesendanger2009spin,Wulfhekel2007}. When tip and sample show a parallel spin polarization, the tunneling conductance is enhanced in comparison to the antiparallel case causing an increase in height. When carefully inspecting the three images recorded under various perpendicular magnetic fields of 0.1 T (upper panel), 0 T (middle panel), and $-$0.1 T (bottom panel), one notices a change in the size of the domains, proving that the observed contrast is of magnetic origin.
Figure~\ref{Fig1}(c) presents the corresponding differential conductance ($dI/dU$) maps of figure~\ref{Fig1}(b). The TMR is much more pronounced in the $dI/dU$ maps (figure~\ref{Fig1}(c)), which directly map the differential conductance. A binary contrast was observed and plotted as dark and bright orange. This contrast reflects the out-of-plane easy axis of the material leading to up and down magnetized domains of about 1 $\mu$m size that are imaged with the out-of-plane sensitive Cr-coated W tips. In analogy to the topographic appearance of the domains, the contrast in differential conductivity more clearly shows the effect of the magnetic field on the domains. Dark domains magnetized up shrink while the bright domains expand when ramping the field to negative values. The spin contrast, however, remains the same between two opposite magnetic field polarities. This indicates that the applied magnetic field neither influences the tip spin-polarization \cite{Menzel2012} nor the sample magnetization axis, i.e.\ the local easy axis of magnetization is perpendicular to the FGT surface. 

\begin{figure*}[ht!]
\includegraphics[width=1\linewidth]{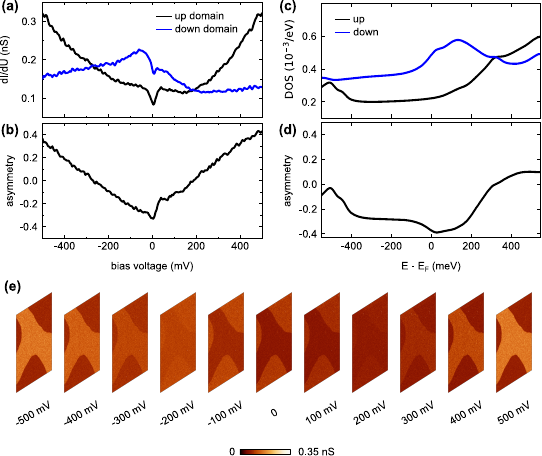}
\caption{(a) Tunneling spectra of spin-up (black) and –down (blue) domains. (b) Asymmetry derived from (a).
(c) Vacuum density of states calculated from DFT for a distance of 4.28 {\AA} above the surface Te atoms. (d) Asymmetry derived from (c).
%\,\mathrm{\AA}
(e) $dI/dU$ maps acquired using spectroscopic imaging from $-$500 to 500 mV with an increment of 100 mV. Image size: 2.3 $\times$ 2.3 $\mu m^2$. Measuring conditions: (a, b) $B$ = $-$0.29 T, $U$ = $-$500 mV, $I$ = 50 pA, and $U_{mod}^{rms}$ = 10 mV, (c) $B$ = 0.15 T, $U$ = $-$500 mV, $I$ = 100 pA, and $U_{mod}^{rms}$ = 40 mV.}
\label{Fig2}
\end{figure*}

To investigate the spin polarization of the electronic states of FGT we used spin-polarized scanning tunneling spectroscopy (SP-STS) on the magnetic domains. To leading order, $dI/dU$ is proportional to the local density of states (LDOS) in the vacuum at the corresponding tip position \cite{Tersoff1985}. Figure~\ref{Fig2}(a) displays the spin-polarized tunneling spectra taken at the spin-up (-down) domain as black (blue) curve. In a central bias voltage range between $\approx$ $-250$ meV and $\approx$ $+200$ meV, the differential conductivity is higher in the down domain compared to the up domain. The corresponding asymmetry (difference of the $dI/dU$ signal for up and down domains divided by their sum) is calculated from figure~\ref{Fig2}(a) and presented in figure~\ref{Fig2}(b) showing a negative asymmetry in the aforementioned bias regime and sign changes at voltages below and above. 
We compare the measured asymmetry to the LDOS calculated from first principles 
%in the vacuum 
which is shown in figure~\ref{Fig2}(c) for the up and down spin channels in black and blue, respectively. The DFT-calculated LDOS is shown in the vacuum above a 3 layer FGT film calculated using the experimental lattice constant \cite{Deiseroth2006} and the local density approximation (LDA) for the exchange-correlation functional (see Methods for details). In agreement with the measured SP-STS we find a minimum in the spin-up LDOS around the Fermi level ($E_\mathrm{F}$) and a peak close to $E_\mathrm{F}$ in the down spin channel. The corresponding asymmetry is shown in figure~\ref{Fig2}(d) which shows a similar overall dependence with bias voltage but is systematically shifted to more negative values.

On both spin-up and spin-down domains, the experimental spectra show a local minimum exactly at the Fermi energy. Such relatively sharp dips in the differential conductivity when tunneling into ferromagnets are often found and are related to inelastic tunneling events in which spin-waves or magnons are excited \cite{Moodera1998,Balashov2006,Gao2008,Balashov2008}. These processes are not considered in the DFT calculations and are thus absent. These inelastic effects will not be further considered in this work.  
Figure~\ref{Fig2}(e) shows the domain contrast in $dI/dU$ at various bias voltages from $-$500 to 500 mV with the increment of 100 mV. In agreement with the asymmetry obtained by STS, the $dI/dU$ contrast reverses at $U = 200$ mV and $U = -250$ mV. 

Complementary to the STM measurements we study FGT using a series of density functional theory (DFT) calculations. Apart from the thin film calculations discussed above, we focus on calculations for bulk FGT crystals. Table~\ref{tab:moments} summarizes the results of calculations with varying parametrizations of the exchange correlation function (generalized gradient (PBE) and local density approximations (LDA)) and changing lattice parameters. In agreement to previously published data \cite{ZHU2021107085}, we find that the average spin moment of the Fe atoms decreases when using the LDA instead of the PBE functional. Upon compression of the FGT lattice, the average Fe moment further decreases down to $1.46\,\mu_\mathrm{B}$ for 5\% compression, which is in better agreement with the experimentally observed values of $1.2-1.8\,\mu_\mathrm{B}$ \cite{Deiseroth2006, doi:10.7566/JPSJ.82.124711}. It is noteworthy that this decrease is in agreement with the typical underestimation of the lattice parameter for the LDA functional and agrees with previously published calculations \cite{ZHU2021107085}.

\begin{table}[htb]
    \centering
    \caption{Average Fe spin moment of bulk Fe$_3$GeTe$_2$ with the corresponding lattice parameters. The theory (DFT) results are given for the PBE and LDA exchange correlation functionals where the same lattice parameter as in Ref.~\cite{Deiseroth2006} was used. The labels ``LDA$^*$'' and ``LDA$^{**}$'' indicate the calculations where the lattice parameters are taken for $3\%$ and $5\%$ isotropic compressive strain, respectively. Experimental values are given for comparison (the numbers in parentheses for Ref.~\cite{May2016} indicate the nominal Fe content per formula unit of two different samples).}
    \begin{tabular}{cl|c|c|c}
        Method & & $a$ (\AA) & $c$ (\AA) & $\braket{\mu_\mathrm{Fe}}$ ($\mu_\mathrm{B}$)\\\hline
        Theory      & PBE      & $3.991$ & $16.333$ & $2.12$ \\
        (this work) & LDA      & $3.991$ & $16.333$ & $1.99$ \\
                    & LDA$^*$  & $3.871$ & $15.840$ & $1.69$ \\
                    & LDA$^{**}$ & $3.791$ & $15.514$ & $1.46$ \\
        Experiment  & Ref.~\cite{Deiseroth2006} & $3.991$ & $16.333$ & $1.20$ \\
                    & Ref.~\cite{doi:10.7566/JPSJ.82.124711} & $3.991$ & $16.296$ & $1.80$ \\
                    & Ref.~\cite{May2016} (2.8) & $3.956$ & $16.395$ & $1.08$ \\
                    & Ref.~\cite{May2016} (3.0) & $4.018$ & $16.334$ & $1.32$
    \end{tabular}
    \label{tab:moments}
\end{table}

\begin{figure}[ht!]
 \includegraphics[width=0.8\linewidth]{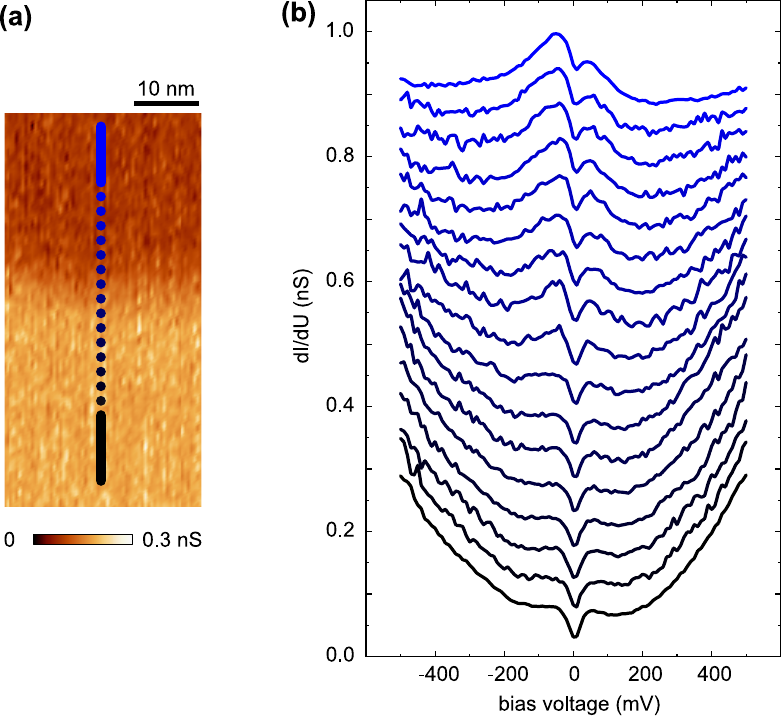}
\caption{(a) $dI/dU$ map of the domain wall. (b) Spectra measured at the location marked by the dotted line in (a). Measuring conditions: $B$ = $-$0.29 T, $U$ = $-$500 mV, $I$ = 50 pA, and $U_{mod}^{rms}$ = 10 mV.}
\label{Fig3}
\end{figure}

In the following, we focus on the structure of the domain wall.
Figure~\ref{Fig3}(a) shows a zoomed image of a magnetic domain boundary, in which the spin contrast smoothly varies between the two saturated signals in the domains. The dotted line marks the positions where the spectra in figure~\ref{Fig3}(b) were acquired. The spectra are vertically offset for clarity. The elongated lines in figure~\ref{Fig3}(a) at both ends of the dotted line show the area for the averaged spectra of the dark (first curve from top) and bright (last curve at the bottom) domain, respectively. In the transition area between the spin-down and spin-up domains, the spectra evolve smoothly from one to the other without the appearance of additional features inside the domain wall. This indicates that while the magnetization smoothly rotates between the two domains, no additional electronic states are created inside the domain wall, i.e.\ there is a smooth transition between the electronic structure with opposite magnetization directions that is trivial without edge states.

\begin{figure}[ht!]
 \includegraphics[width=0.5\linewidth]{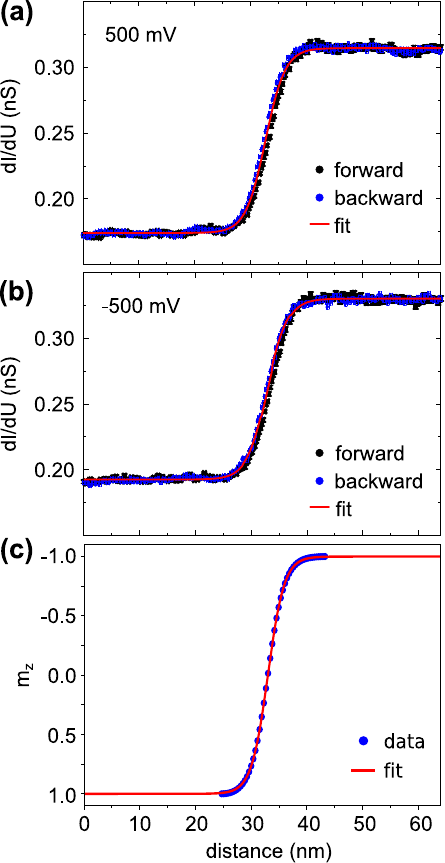}
\caption{Averaged dI/dU line scans of the domain wall at zero magnetic field at voltages of (a) 500 mV and (b) $-$500 mV. Black and blue data corresponds to forward and backward scans respectively. The red curves present the fitted results. Measuring conditions: $I$ = 100 pA, and $U_{mod}^{rms}$ = 100 mV. (c) Domain-wall profile from atomistic spin-dynamics simulations using DFT-calculated exchange parameters.
}
\label{Fig4}
\end{figure}

The relatively low noise in the differential conductance and the use of an antiferromagnetic tip allows to precisely measure the wall profile of the domain wall.
Figure~\ref{Fig4}(a) and \ref{Fig4}(b) show the fully resolved wall profiles at $U$ = 500 mV and $-$500 mV, respectively. The forward and backward scans (black and blue curves, respectively) show no significant difference highlighting the absence of magnetic interaction between the tip and the wall that potentially could drag the wall during scanning across it.
The averaged profile was fitted using a hyperbolic tangent function (red curves), which corresponds to the micromagnetic description of a domain-wall profile for Bloch walls in bulk materials with uniaxial anisotropy \cite{hubert-schafer2000}, 
\begin{equation}
    \frac{\mathrm d I}{\mathrm d U} \left( x \right) \propto \tanh{\frac{2x}{w}},
    \label{eq:dw_profile}
\end{equation}
\noindent where $x$ corresponds to the profile distance and $w$ the domain-wall width. The fits nicely reproduce the shape of the domain wall. This indicates that at the surface, the wall profile still follows that of a conventional Bloch wall and that no N\'eel-caps are present at the surface. The latter would result in asymmetric wall profiles that significantly deviate from the simple analytic wall shape \cite{Scheinfein1991,Schlickum2003}.
Furthermore, the domain-wall width $w$ contains information about the exchange stiffness $A$ and the uniaxial anisotropy constant $K$, which can be expressed as,
\begin{equation}
    w = 2\sqrt{\left|\frac{A}{K}\right|}. 
    \label{eq:dw_width}
\end{equation}
From the fitted curves we extract a domain width of $7.11 \pm 0.03$ nm and $7.32 \pm 0.02$ nm for $U$ = 500 mV and $-$500 mV, respectively.

Based on our DFT calculations for bulk FGT we compute parameters for the extended classical Heisenberg Hamiltonian describing the interaction of spins $\vec{S}_{i}=\vec{M}_i/\mu_{i}$ ($\mu_i=\lvert\vec{M}_i\rvert$) sitting at lattice sites $i$:
\begin{eqnarray}
    H = &-&\sum_{\braket{ij}}J_{ij} \left( \vec{S}_i \cdot \vec{S}_j \right) -\sum_{\braket{ij}}\vec{D}_{ij} \cdot \left( \vec{S}_i \times \vec{S}_j \right)\nonumber \\
    &-& \sum_{i} K (\vec{S}_i \cdot \hat{\vec{e}}_z)^2 - \sum_{i} \mu_{i} \vec{B}\cdot \vec{S}_{i} \nonumber \\
    &-& \frac{\mu_{0}}{8\pi}\sum_{i\neq j} \frac{3\left(\vec{S}_i \cdot \hat{\vec{r}}_{ij}\right)\left(\vec{S}_j \cdot \hat{\vec{r}}_{ij}\right)-\vec{S}_i\cdot\vec{S}_j}{r_{ij}^3} \ ,
    \label{eq:Heisenberg}
\end{eqnarray} 
where $J_{ij}$ and $\vec{D}_{ij}$ describe the isotropic and antisymmetric exchange interactions (also referred to as Dzyaloshinskii-Moriya interaction). The third and fourth terms contain the contributions arising from the on-site anisotropy $K$, the Zeeman energy due to an external magnetic field $\vec{B}$, as well as the dipolar contribution.

\begin{figure*}[ht!]
 \includegraphics[width=1\linewidth]{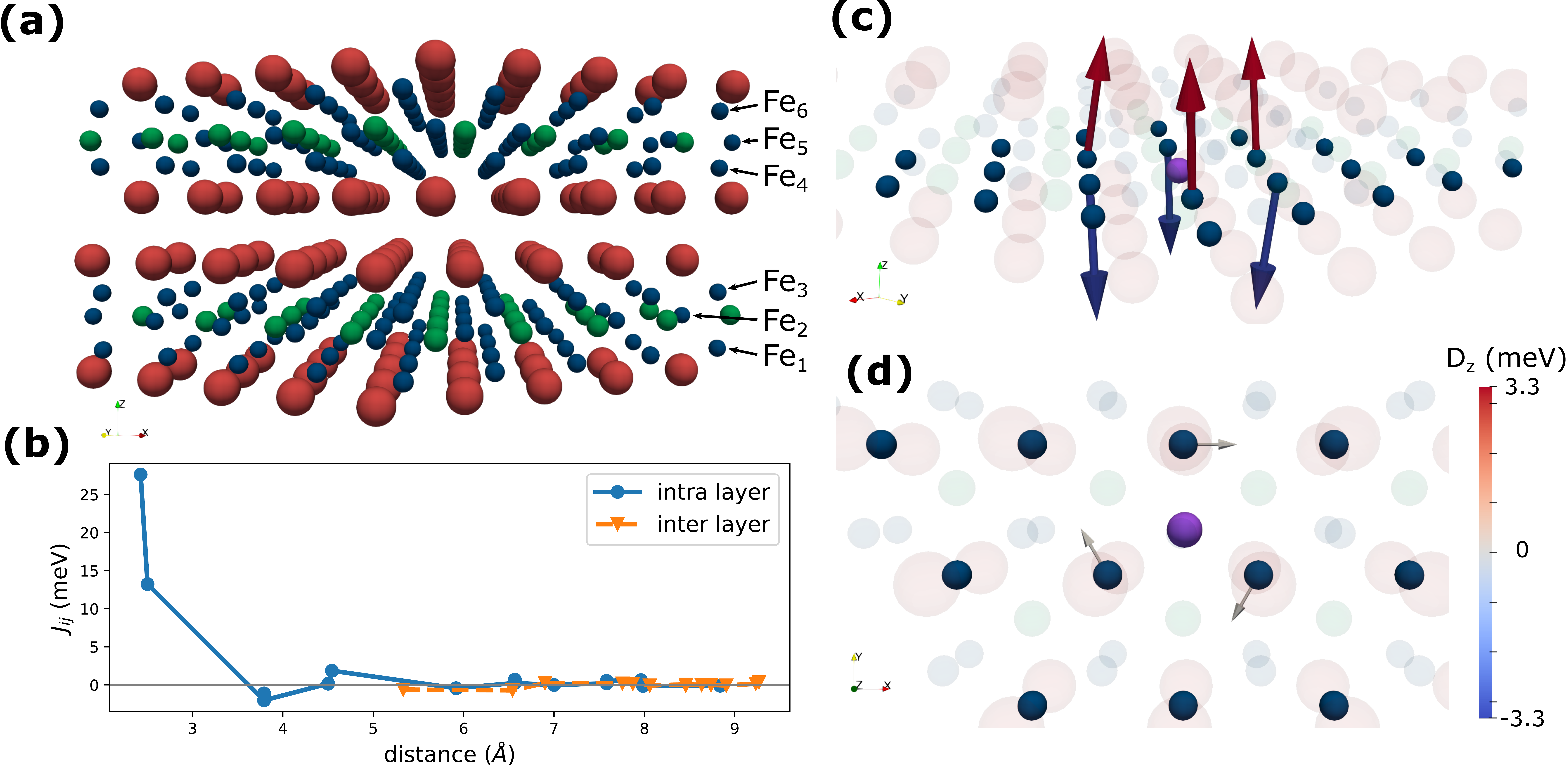}
\caption{(a) Crystal structure of Fe$_3$GeTe$_2$ where the six different Fe layers in the unit cell are marked with labels ``Fe$_1$'' to ``Fe$_6$''. (b) Exchange coupling constants $J_{ij}$ vs.\ distance between the Fe atoms (intra-layer coupling only between Fe$_1$, Fe$_2$ and Fe$_3$, inter-layer coupling for $i\in[1,2,3]$ and $j\in[4,5,6]$). (c) $\vec{D}_{ij}$ for pairs of Fe$_1$-Fe$_1$ and (d) for Fe$_1$-Fe$_2$ where the DMI vector is shown at the position of atom $j$ and atom $i$ is highlighted by the purple sphere.}
\label{Fig6}
\end{figure*}

In figure~\ref{Fig6}(a) we show the alternating layers of the FGT lattice where the six different Fe atom layers are labelled Fe$_1$ to Fe$_6$. The calculated exchange interactions ($J_{ij}$) are shown as a function of the distance between the Fe atoms in figure~\ref{Fig6}(b). The data is shown for the smallest lattice constant (5\% compressed compared to the experimental lattice constant) where the computed average Fe moment is smallest and agrees best with the experiment (cf.\ Tab.~\ref{tab:moments}). The large spread in the experimental values of the average magnetic moment could be the result of varying Fe content and increasing $c/a$ ratio in different samples \cite{May2016}. We find strong ferromagnetic inter-layer couplings at small distances which decay with additional sign changes to antiferromagnetic ($J_{ij}<0$) coupling at larger Fe-Fe distances. The Dzyaloshinskii-Moriya interactions (DMI) are only significant for inter-layer couplings. As shown in figure~\ref{Fig6}(c), the largest DMI is found for nearest neighbor Fe$_1$-Fe$_1$ pairs, with vectors mainly pointing in the direction perpendicular to the FGT film and alternating in the sign for adjacent neighbors. The pairs in the Fe$_3$-Fe$_3$ layer are related by mirror symmetry and the DMI therefore exactly cancels the contributions from Fe$_1$-Fe$_1$. The second-largest contribution comes from nearest neighbors between Fe$_1$ and Fe$_2$ atoms (or the symmetry equivalent Fe$_3$-Fe$_2$ pairs) as shown in figure~\ref{Fig6}(d). All other DMI pairs are smaller by at least an order of magnitude.

The micromagnetic spin stiffness and spiralization vectors
\begin{eqnarray}
    A &=& \frac{1}{2V}\sum_{j>0} R_{0j}^2J_{0j} \ , \\
    \vec{D} &=& \frac{1}{V}\sum_{j>0} R_{0j}\vec{D}_{0j}\ ,
\end{eqnarray}
where $V$ is the volume per Fe atom, can be calculated from the atomistic exchange parameters \cite{Schweflinghaus2016}. These parameters define the micromagnetic energy functional 
\begin{equation}
    E[\vec{m}] = \int \mathrm{d}\vec{r} \Bigl[ A\bigl(\dot{\vec{m}}\bigr)^2 + \vec{D} \cdot \bigl(\vec{m} \times \dot{\vec{m}}\bigr) + K m_z^2 \Bigr]
\end{equation}
for a continuous magnetization field $\vec{m}[\vec{r}]$ with $\dot{\vec{m}} = \nabla \vec{m}[\vec{r}]$.
The crystal symmetries of the FGT layers and the structure of the calculated DMI vectors lead to vanishing spiralization $\vec{D}$. The computed spin stiffness $A$, the anisotropy $K$ and the resulting micromagnetic domain-wall width (eq.~\ref{eq:dw_width}) are summarized in Tab.~\ref{tab:stiffness_etc}. The parameters are shown for three calculations using the experimental lattice constant as well as for 3\% and 5\% smaller lattice parameters. We can see that the spin stiffness as well as the anisotropy get smaller with smaller lattice constant which matches the observed decrease in the average Fe moment. The micromagnetic domain-wall width on the other hand increases from $5$ to $6.3$\,nm which is close to the measured domain-wall width of around $7$\,nm.

We further use the atomistic exchange parameters in spin-dynamics simulations. We consider a $50\times5\times2$ supercell and find the magnetic ground state starting from random spin orientations (see Methods for details). For the experimental lattice constant and the calculation with 3\% compressive strain we find antiferromagnetically coupled FGT layers, in contrast to experimental findings but in agreement with previous theoretical results \cite{ZHU2021107085}. Only when the FGT lattice is compressed by 5\% we find ferromagnetically coupled FGT layers. We furthermore calculate the domain-wall width by pinning the left and right end of the supercell to point in $\pm z$ direction (i.e.\ parallel to the surface normal of the FGT layers). This creates a domain wall in the center of the supercell. We fit the resulting domain-wall profile (shown in figure~\ref{Fig4}(c)) using eq.~(\ref{eq:dw_profile}). The fitted domain-wall width is shown in Tab.~\ref{tab:stiffness_etc} where we observe that the atomistic domain-wall width agrees well with the micromagnetic parameters and in particular also the experiment.

\begin{table}[t!]
    \centering
    \caption{Exchange stiffness ($A$), anisotropy ($K$) and resulting micromagnetic ($w_\mathrm{micro}=2\sqrt{|A/K|}$) and atomistic ($w_\mathrm{atom}$) domain-wall widths for bulk Fe$_3$GeTe$_2$ with different lattice parameters (cf.\ Tab.~\ref{tab:moments}). The FM/AFM label indicates whether antiferromagnetic or ferromagnetic ordering is found between the FGT layers. $w_\mathrm{atom}$ is the result of atomistic spin dynamics simulations using the Spirit code with the respective set of exchange coupling parameters.}
    \begin{tabular}{c|c|c|c|c|c|c}
Method   & $A$ (meV\,nm/Fe) & $A$ (pJ/m) & $K$ (meV/Fe) & $K$ (MJ/m$^3$) & $w_\mathrm{micro}$ (nm) & $w_\mathrm{atom}$ (nm) \\ \hline
LDA (AFM) & $3.248$ & $13.859$ & $-0.506$ & $-2.16$ & $5.067$ & $5.489$ \\
LDA$^*$ (AFM) & $2.403$ & $11.237$ & $-0.263$ & $-1.23$ & $6.046$ & $6.094$ \\
LDA$^{**}$ (FM) &  $1.772$ & $8.818$ & $-0.180$ & $-0.897$ & $6.274$ & $5.927$
    \end{tabular}
    \label{tab:stiffness_etc}
\end{table}

Note that the slight difference in the domain wall as a function of the sign of bias voltage found in the experiment is larger than the experimental error of the fitted domain-wall width, i.e.\ it seems to be significant. We have carried out a more systematic study of this difference in more than 50 individual experiments. We find a 5$\pm$1\% wider wall width at negative sample bias. A possible explanation for this weak effect is the modification of the magnetic anisotropy by the electric field in the tunneling junction, i.e.\ magneto-electric coupling (MEC) \cite{smolenskiui1982ferroelectromagnets, lottermoser2004magnetic, spaldin2005renaissance, zheng2004multiferroic, zavaliche2005electric, duan2006predicted, Weisheit2007, Duan2008, fechner2008magnetic, maruyama2009large, subkow2011electron, shiota2012induction, tournerie2012influence, nan2014, gerhard2014influence}. MEC was suggested to be present in FGT  modifying the anisotropy or the exchange \cite{Deng2018,Liu2019}, as also seen in STM experiments from variations of the tunneling spectra with tip-sample distance \cite{Zhao2021}. Besides a direct MEC, the domain-wall width can also be influenced by an indirect effect, in which the electric field in the junction leads to changes in the relaxation, i.e.\ vertical lattice constant \cite{Gerhard2010}. As our DFT calculations indicate, both the anisotropy and the exchange depend sensitively on the interlayer distance. The latter is susceptible to the electric field as the layered structure of FGT with its local charges strongly couples to the field.  
The observed 5\% change of the domain-wall width would correspond to a twice as large change in the ratio of the exchange constant and the uniaxial anisotropy by the electric field in the STM junction highlighting the high susceptibility of FGT on the application of an electric field.

\begin{figure*}[ht!]
 \includegraphics[width=1\linewidth]{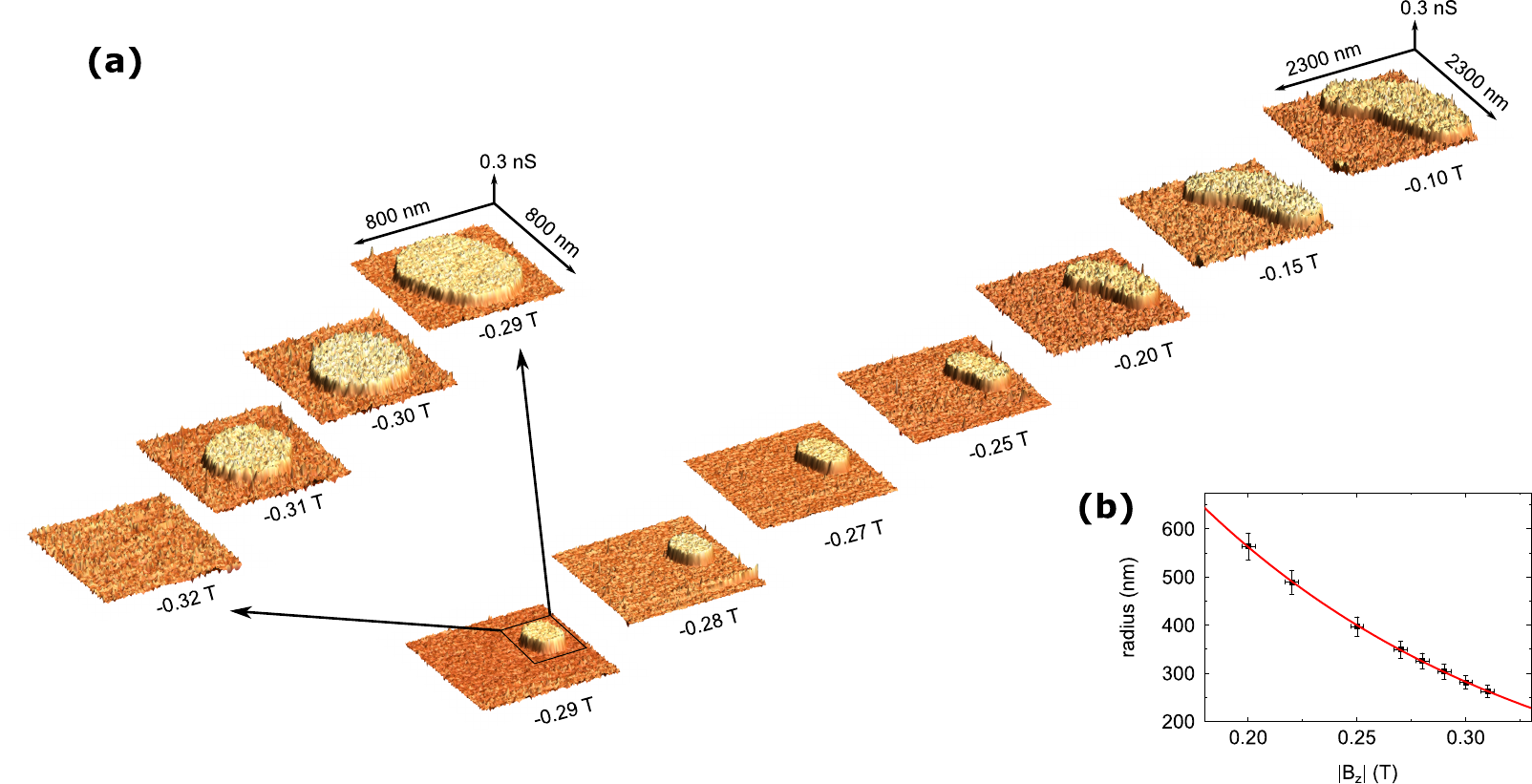}
\caption{(a) Pseudo three-dimensional representation of the $dI/dU$ maps under various of magnetic fields. (b) Radius of the skyrmionic bubble as a function of magnetic fields. Measuring conditions: $U$ = $-$500 mV, $I$ = 50 pA, and $U_{mod}^{rms}$ = 80 mV.}
\label{Fig5}
\end{figure*}

The out-of-plane domain-wall with their 180$^\circ$ domain walls essentially are bubble domains \cite{ding2019observation}, where the sense of rotation in the wall is kept along the wall. A change of the sense of rotation would lead to singularities in form of Bloch lines and cusps in the wall \cite{han2019topological}, which we do not observe, i.e., the sense of rotation of the 180$^\circ$ domain wall between the central domain of the bubble and its surrounding does not change within the domain wall of the bubble. In this respect, the bubble domains are topologically equivalent to skyrmions, i.e.\ carry a topological charge of the winding of the magnetization and are topologically protected from transitions to the trivial, i.e.\ single domain ferromagnetic state. Note that with SP-STM, we, however, measure only one component of the magnetization (z-direction), such that we cannot determine the sense of rotation of the wall.
To investigate the topological protection of the bubble domains, we investigated their evolution in large out-of-plane magnetic fields, analogous to Lorentz transmission electron microscopy studies of thin FGT flakes by Ding et al. \cite{ding2019observation}. Figure \ref{Fig5}(a) displays the evolution of a single and several $\mu$m sized bubble domain in a magnetic field that compresses the bubble. With rising magnetic field, the irregularly shaped bubble domain is compressed to a circular shape of sub $\mu$m size ($-$0.3 T) followed by a further compression \cite{Romming2015,Rohart2016,bernand2018skyrmion} until it suddenly collapses between 0.31 T and 0.32 T. This collapse annihilates the topological charge of the bubble domain and is associates with an energetic barrier \cite{Leonov2015a,Rohart2016,mougel2020instability}.
Note that all bubbles collapsed roughly at the same critical field, which is also reflected by the fact that the overall magnetization curve (not shown) along the easy magnetic direction saturates at $\approx 0.32$ T.
Simple continuum theory predicts that the lateral size of a topologically protected skyrmion should be proportional to $1/B$ \cite{Bogdanov1994a,Wilson2014}.
Figure \ref{Fig5}(b) displays the effective radius $r_\mathrm{eff}$ of the bubble ($r_\mathrm{eff}=\sqrt{\mathrm{area}/\pi}$) versus magnetic field. 
The effective radius of the bubble monotonically decreases with magnetic fields to around 262 nm before it vanishes. This underlines that the relatively large bubbles can be fully described by a continuum model. They are compressed but not to atomic sizes before the collapse, i.e.\ they become energetically unstable before shrinking to a singularity. They, however, do not follow the simple $1/B$ behavior but shrink slower than that at high fields. Following the analytical description of the phase diagram of bubbles and skyrmions by Bernand-Mantel et al. \cite{bernand2018skyrmion} and considering that the DMI in the FGT double layers cancels (or almost cancels), the collapse of the bubbles in a bulk sample is predicted, when the Zeeman energy, which destabilizes the bubble, exceeds the long-range dipolar energy, which stabilizes domains in the film. This is expected when the external magnetic field approaches the demagnetization field. Equation 9 of that paper predicts bubble collapse for bulk samples (radius $r$ of the bubble $<<$ sample thickness $t$) to happen for $H=\frac{2M_s}{\pi}$ corresponding to  $B=$ 0.3 T in very good agreement with the experimental observation. This independently confirms that the DMI in FGT is small compared to the other interactions. 

\section*{Conclusion}

In conclusion, we determined the spin-polarized density of states of FGT using SP-STM and were able to resolve the internal structure of the 180$^\circ$ Bloch walls of the material. It agrees well with simulations based on DFT parameters. The domain-wall width, however, was found to depend slightly on the electric field indicating a local magneto-electric effect in the tunneling junction of the STM. Finally, we followed the size evolution of magnetic bubble domains down to their collapse at about 0.32 T.

\section*{METHODS}

\subsection{FGT crystal growth and chemical analysis}

High-quality Fe$_{3-\delta}$GeTe$_2$ $(\delta \approx 0.04)$ single-crystals were grown by chemical vapour transport method using I$_2$ and TeCl$_4$ as transport agent. Pure elements of Fe (Alfa Aesar, 99.999\%), Ge (Merck, 99.999\%), and Te (G-Materials, 99.999\%) were mixed in the molar ratio of 3.1:1:2, respectively. The chemicals were filled in a fused silica tube that was evacuated and sealed, subsequently. The ampule was placed in a zone-furnace with a temperature gradient of 750\degree C - 670\degree C for 10 days. Single-crystals with a hexagonal sheet-like morphology were grown at the low-temperature end of the ampule.
The Chemical composition of the crystals was examined by energy-dispersive x-ray spectroscopy employing a Coxem EM-30$^N$ benchtop scanning electron microscope equipped with an Oxford Instruments detection system. The Fe deficiency $\delta$ from the stoichiometric composition of FGT was determined to 4\%. We also determined the magnetocrystalline anisotropy from hard axis magnetization loops of the samples to 1.46$\times 10^6$ J/m$^3$.

\subsection{SP-STM measurements}

The FGT crystals were \textit{in-situ} exfoliated under the base pressure of $ \approx 1 \times 10^{-9}$ mbar. For this, one end of UHV compatible Kapton tape was attached to the bulk FGT sample, the other on the sample rack of the load lock. When the load lock was pumped down to base pressure, the sample was taken out of the sample rack with a wobble stick and hence the sample surface was exfoliation at the same time, thinning the bulk sample and forming a fresh and flat surface for the STM investigation. The SP-STM measurements were performed using a home-build low-temperature (0.7 K) STM and a Nanonis control system with ultra-high vacuum (UHV) of $\approx 1 \times 10^{-10}$ mbar. The STM equipped a superconducting magnet, which can generate magnetic fields perpendicular to the sample surface up to 4 T. The spin-polarized tip was prepared by depositing $\approx 50$ ML Cr on an electrochemically etched W tip that had been pre-annealed in UHV. The tunneling spectra and images were recorded in a standard lock-in method with the root-mean-square bias voltage modulation ($U_{mosd}^{rms}$) of 10 or 40 or 80 mV at 1.08 kHz. All images were processed using Nanotech WSxM \cite{horcas2007}.

\subsection{DFT and spin-dynamics calculations}

We performed density functional theory (DFT) calculations for a thin film of Fe$_2$GeTe$_3$ consisting of three FGT layers using the experimental lattice constant \cite{Villars2016:sm_isp_sd_1420956}. We perform our calculations with the \code{JuKKR} code \cite{jukkr} that implements the full-potential relativistic Korringa-Kohn-Rostoker Green function method (KKR) \cite{Ebert2011} with an exact description of the shape of the atomic cell \cite{Stefanou1990, Stefanou1991}. We use an angular momentum cutoff of $\ell_\mathrm{max}=3$ and the local density  (LDA) and generalized gradient approximations (GGA) to the exchange-correlation potential \cite{Vosko1980, PBE}. Our calculations of the exchange coupling constants for the extended Heisenberg Hamiltonian are based on the method of infinitesimal rotations \cite{Liechtenstein1987, PhysRevB.68.104436}.

The Heisenberg exchange coupling parameters are then used in spin-dynamics simulations using the \code{Spirit} code \cite{spirit-paper, spirit-code}. 
Our atomistic spin-dynamics simulations are based on the extended Heisenberg Hamiltonian described in the text. We use a $50\times5\times2$ supercell with periodic boundary conditions in all directions. Note that each unit cell consists of six magnetic Fe atoms. We verified that this size is sufficient by calculating a larger supercell of length 100 instead of 50 which reproduces the domain wall width within 1\%. The spin dynamics calculations use the time-evolution of the direction of the spins based on the Landau-Lifshitz-Gilbert (LLG) equation which we stop if the forces on the spins converge to $10^{-7}$ or after 200k time steps.

The series of DFT and spin-dynamics calculations of this work are orchestrated with the help of the AiiDA-KKR and AiiDA-Spirit plugins \cite{aiida-kkr, aiida-spirit} that connect the \code{JuKKR} and \code{Spirit} codes to the AiiDA environment \cite{aiida}.

\section*{ACKNOWLEDGEMENTS}

This work was funded by the Deutsche Forschungsgemeinschaft (DFG, German Research Foundation) under Germany's Excellence Strategy – Cluster of Excellence Matter and Light for Quantum Computing (ML4Q) EXC 2004/1 – 390534769 and TRR 288 – 422213477 (project B06 and B03) and CRC 1238 -- 277146847 (Project C01). We acknowledge computing time granted by the JARA Vergabegremium and provided on the JARA Partition part of the supercomputer CLAIX at RWTH Aachen University as well as support by the Alexander-von-Humboldt Foundation and the Innovationspool project "Solid-State Quantum Computing" of the Helmholtz Association. We thank Betrand Dup\'e for the discussions.

\section*{AUTHOR CONTRIBUTIONS}
H.-H.\ Y., N.\ B.\ and W.\ W.\ carried out STM experiments, A.-A.\ H., \ K.\ S.\ and M.\ L.\ T\ grew the samples and carried out the characterization of the bulk material.  P.\ R.\ and D.\ G.\ performed DFT calculations.  P.\ R., D.\ G., S.\ B., M.\ H., L.\ Z.\ and Y.\ M.\ analyzed the DFT calculations and the resulting exchange coupling constants, M.\ H.\ and P.\ R.\ ran the atomistic spin-dynamics simulations. Q.\ L.\ performed additional micromagnetic calculations. Y.\ M.\ and W.\ W.\ conceived the study. All authors contributed in the discussion of the results and to writing the manuscript. 

%\bibliography{FGT}

\begin{thebibliography}{74}%
\makeatletter
\providecommand \@ifxundefined [1]{%
 \@ifx{#1\undefined}
}%
\providecommand \@ifnum [1]{%
 \ifnum #1\expandafter \@firstoftwo
 \else \expandafter \@secondoftwo
 \fi
}%
\providecommand \@ifx [1]{%
 \ifx #1\expandafter \@firstoftwo
 \else \expandafter \@secondoftwo
 \fi
}%
\providecommand \natexlab [1]{#1}%
\providecommand \enquote  [1]{``#1''}%
\providecommand \bibnamefont  [1]{#1}%
\providecommand \bibfnamefont [1]{#1}%
\providecommand \citenamefont [1]{#1}%
\providecommand \href@noop [0]{\@secondoftwo}%
\providecommand \href [0]{\begingroup \@sanitize@url \@href}%
\providecommand \@href[1]{\@@startlink{#1}\@@href}%
\providecommand \@@href[1]{\endgroup#1\@@endlink}%
\providecommand \@sanitize@url [0]{\catcode `\\12\catcode `\$12\catcode
  `\&12\catcode `\#12\catcode `\^12\catcode `\_12\catcode `\%12\relax}%
\providecommand \@@startlink[1]{}%
\providecommand \@@endlink[0]{}%
\providecommand \url  [0]{\begingroup\@sanitize@url \@url }%
\providecommand \@url [1]{\endgroup\@href {#1}{\urlprefix }}%
\providecommand \urlprefix  [0]{URL }%
\providecommand \Eprint [0]{\href }%
\providecommand \doibase [0]{http://dx.doi.org/}%
\providecommand \selectlanguage [0]{\@gobble}%
\providecommand \bibinfo  [0]{\@secondoftwo}%
\providecommand \bibfield  [0]{\@secondoftwo}%
\providecommand \translation [1]{[#1]}%
\providecommand \BibitemOpen [0]{}%
\providecommand \bibitemStop [0]{}%
\providecommand \bibitemNoStop [0]{.\EOS\space}%
\providecommand \EOS [0]{\spacefactor3000\relax}%
\providecommand \BibitemShut  [1]{\csname bibitem#1\endcsname}%
\let\auto@bib@innerbib\@empty
%</preamble>
\bibitem [{\citenamefont {Ma}\ \emph {et~al.}(2012)\citenamefont {Ma},
  \citenamefont {Dai}, \citenamefont {Guo}, \citenamefont {Niu}, \citenamefont
  {Zhu},\ and\ \citenamefont {Huang}}]{ma2012evidence}%
  \BibitemOpen
  \bibfield  {author} {\bibinfo {author} {\bibfnamefont {Y.}~\bibnamefont
  {Ma}}, \bibinfo {author} {\bibfnamefont {Y.}~\bibnamefont {Dai}}, \bibinfo
  {author} {\bibfnamefont {M.}~\bibnamefont {Guo}}, \bibinfo {author}
  {\bibfnamefont {C.}~\bibnamefont {Niu}}, \bibinfo {author} {\bibfnamefont
  {Y.}~\bibnamefont {Zhu}}, \ and\ \bibinfo {author} {\bibfnamefont
  {B.}~\bibnamefont {Huang}},\ }\href@noop {} {\bibfield  {journal} {\bibinfo
  {journal} {ACS Nano}\ }\textbf {\bibinfo {volume} {6}},\ \bibinfo {pages}
  {1695} (\bibinfo {year} {2012})}\BibitemShut {NoStop}%
\bibitem [{\citenamefont {Leb{\`e}gue}\ \emph {et~al.}(2013)\citenamefont
  {Leb{\`e}gue}, \citenamefont {Bj{\"o}rkman}, \citenamefont {Klintenberg},
  \citenamefont {Nieminen},\ and\ \citenamefont {Eriksson}}]{lebegue2013two}%
  \BibitemOpen
  \bibfield  {author} {\bibinfo {author} {\bibfnamefont {S.}~\bibnamefont
  {Leb{\`e}gue}}, \bibinfo {author} {\bibfnamefont {T.}~\bibnamefont
  {Bj{\"o}rkman}}, \bibinfo {author} {\bibfnamefont {M.}~\bibnamefont
  {Klintenberg}}, \bibinfo {author} {\bibfnamefont {R.~M.}\ \bibnamefont
  {Nieminen}}, \ and\ \bibinfo {author} {\bibfnamefont {O.}~\bibnamefont
  {Eriksson}},\ }\href@noop {} {\bibfield  {journal} {\bibinfo  {journal}
  {Phys. Rev. X}\ }\textbf {\bibinfo {volume} {3}},\ \bibinfo {pages} {031002}
  (\bibinfo {year} {2013})}\BibitemShut {NoStop}%
\bibitem [{\citenamefont {Tong}\ \emph {et~al.}(2016)\citenamefont {Tong},
  \citenamefont {Gong}, \citenamefont {Wan},\ and\ \citenamefont
  {Duan}}]{Tong2016}%
  \BibitemOpen
  \bibfield  {author} {\bibinfo {author} {\bibfnamefont {W.-Y.}\ \bibnamefont
  {Tong}}, \bibinfo {author} {\bibfnamefont {S.-J.}\ \bibnamefont {Gong}},
  \bibinfo {author} {\bibfnamefont {X.}~\bibnamefont {Wan}}, \ and\ \bibinfo
  {author} {\bibfnamefont {C.-G.}\ \bibnamefont {Duan}},\ }\href {\doibase
  10.1038/ncomms13612} {\bibfield  {journal} {\bibinfo  {journal} {Nat.
  Commun.}\ }\textbf {\bibinfo {volume} {7}},\ \bibinfo {pages} {13612}
  (\bibinfo {year} {2016})}\BibitemShut {NoStop}%
\bibitem [{\citenamefont {Huang}\ \emph {et~al.}(2017)\citenamefont {Huang},
  \citenamefont {Clark}, \citenamefont {Navarro-Moratalla}, \citenamefont
  {Klein}, \citenamefont {Cheng}, \citenamefont {Seyler}, \citenamefont
  {Zhong}, \citenamefont {Schmidgall}, \citenamefont {McGuire}, \citenamefont
  {Cobden}, \citenamefont {Yao}, \citenamefont {Xiao}, \citenamefont
  {Jarillo-Herrero},\ and\ \citenamefont {Xu}}]{Huang2017}%
  \BibitemOpen
  \bibfield  {author} {\bibinfo {author} {\bibfnamefont {B.}~\bibnamefont
  {Huang}}, \bibinfo {author} {\bibfnamefont {G.}~\bibnamefont {Clark}},
  \bibinfo {author} {\bibfnamefont {E.}~\bibnamefont {Navarro-Moratalla}},
  \bibinfo {author} {\bibfnamefont {D.~R.}\ \bibnamefont {Klein}}, \bibinfo
  {author} {\bibfnamefont {R.}~\bibnamefont {Cheng}}, \bibinfo {author}
  {\bibfnamefont {K.~L.}\ \bibnamefont {Seyler}}, \bibinfo {author}
  {\bibfnamefont {D.}~\bibnamefont {Zhong}}, \bibinfo {author} {\bibfnamefont
  {E.}~\bibnamefont {Schmidgall}}, \bibinfo {author} {\bibfnamefont {M.~A.}\
  \bibnamefont {McGuire}}, \bibinfo {author} {\bibfnamefont {D.~H.}\
  \bibnamefont {Cobden}}, \bibinfo {author} {\bibfnamefont {W.}~\bibnamefont
  {Yao}}, \bibinfo {author} {\bibfnamefont {D.}~\bibnamefont {Xiao}}, \bibinfo
  {author} {\bibfnamefont {P.}~\bibnamefont {Jarillo-Herrero}}, \ and\ \bibinfo
  {author} {\bibfnamefont {X.}~\bibnamefont {Xu}},\ }\href {\doibase
  10.1038/nature22391} {\bibfield  {journal} {\bibinfo  {journal} {Nature}\
  }\textbf {\bibinfo {volume} {546}},\ \bibinfo {pages} {270} (\bibinfo {year}
  {2017})}\BibitemShut {NoStop}%
\bibitem [{\citenamefont {Jiang}\ \emph {et~al.}(2018)\citenamefont {Jiang},
  \citenamefont {Li}, \citenamefont {Wang}, \citenamefont {Mak},\ and\
  \citenamefont {Shan}}]{Jiang2018}%
  \BibitemOpen
  \bibfield  {author} {\bibinfo {author} {\bibfnamefont {S.}~\bibnamefont
  {Jiang}}, \bibinfo {author} {\bibfnamefont {L.}~\bibnamefont {Li}}, \bibinfo
  {author} {\bibfnamefont {Z.}~\bibnamefont {Wang}}, \bibinfo {author}
  {\bibfnamefont {K.~F.}\ \bibnamefont {Mak}}, \ and\ \bibinfo {author}
  {\bibfnamefont {J.}~\bibnamefont {Shan}},\ }\href {\doibase
  10.1038/s41565-018-0135-x} {\bibfield  {journal} {\bibinfo  {journal} {Nat.
  Nanotechnol.}\ }\textbf {\bibinfo {volume} {13}},\ \bibinfo {pages} {549}
  (\bibinfo {year} {2018})}\BibitemShut {NoStop}%
\bibitem [{\citenamefont {Gong}\ \emph {et~al.}(2017)\citenamefont {Gong},
  \citenamefont {Li}, \citenamefont {Li}, \citenamefont {Ji}, \citenamefont
  {Stern}, \citenamefont {Xia}, \citenamefont {Cao}, \citenamefont {Bao},
  \citenamefont {Wang}, \citenamefont {Wang}, \citenamefont {Qiu},
  \citenamefont {Cava}, \citenamefont {Louie}, \citenamefont {Xia},\ and\
  \citenamefont {Zhang}}]{Gong2017}%
  \BibitemOpen
  \bibfield  {author} {\bibinfo {author} {\bibfnamefont {C.}~\bibnamefont
  {Gong}}, \bibinfo {author} {\bibfnamefont {L.}~\bibnamefont {Li}}, \bibinfo
  {author} {\bibfnamefont {Z.}~\bibnamefont {Li}}, \bibinfo {author}
  {\bibfnamefont {H.}~\bibnamefont {Ji}}, \bibinfo {author} {\bibfnamefont
  {A.}~\bibnamefont {Stern}}, \bibinfo {author} {\bibfnamefont
  {Y.}~\bibnamefont {Xia}}, \bibinfo {author} {\bibfnamefont {T.}~\bibnamefont
  {Cao}}, \bibinfo {author} {\bibfnamefont {W.}~\bibnamefont {Bao}}, \bibinfo
  {author} {\bibfnamefont {C.}~\bibnamefont {Wang}}, \bibinfo {author}
  {\bibfnamefont {Y.}~\bibnamefont {Wang}}, \bibinfo {author} {\bibfnamefont
  {Z.~Q.}\ \bibnamefont {Qiu}}, \bibinfo {author} {\bibfnamefont {R.~J.}\
  \bibnamefont {Cava}}, \bibinfo {author} {\bibfnamefont {S.~G.}\ \bibnamefont
  {Louie}}, \bibinfo {author} {\bibfnamefont {J.}~\bibnamefont {Xia}}, \ and\
  \bibinfo {author} {\bibfnamefont {X.}~\bibnamefont {Zhang}},\ }\href
  {\doibase 10.1038/nature22060} {\bibfield  {journal} {\bibinfo  {journal}
  {Nature}\ }\textbf {\bibinfo {volume} {546}},\ \bibinfo {pages} {265}
  (\bibinfo {year} {2017})}\BibitemShut {NoStop}%
\bibitem [{\citenamefont {Deng}\ \emph {et~al.}(2018)\citenamefont {Deng},
  \citenamefont {Yu}, \citenamefont {Song}, \citenamefont {Zhang},
  \citenamefont {Wang}, \citenamefont {Sun}, \citenamefont {Yi}, \citenamefont
  {Wu}, \citenamefont {Wu}, \citenamefont {Zhu}, \citenamefont {Wang},
  \citenamefont {Chen},\ and\ \citenamefont {Zhang}}]{Deng2018}%
  \BibitemOpen
  \bibfield  {author} {\bibinfo {author} {\bibfnamefont {Y.}~\bibnamefont
  {Deng}}, \bibinfo {author} {\bibfnamefont {Y.}~\bibnamefont {Yu}}, \bibinfo
  {author} {\bibfnamefont {Y.}~\bibnamefont {Song}}, \bibinfo {author}
  {\bibfnamefont {J.}~\bibnamefont {Zhang}}, \bibinfo {author} {\bibfnamefont
  {N.~Z.}\ \bibnamefont {Wang}}, \bibinfo {author} {\bibfnamefont
  {Z.}~\bibnamefont {Sun}}, \bibinfo {author} {\bibfnamefont {Y.}~\bibnamefont
  {Yi}}, \bibinfo {author} {\bibfnamefont {Y.~Z.}\ \bibnamefont {Wu}}, \bibinfo
  {author} {\bibfnamefont {S.}~\bibnamefont {Wu}}, \bibinfo {author}
  {\bibfnamefont {J.}~\bibnamefont {Zhu}}, \bibinfo {author} {\bibfnamefont
  {J.}~\bibnamefont {Wang}}, \bibinfo {author} {\bibfnamefont {X.~H.}\
  \bibnamefont {Chen}}, \ and\ \bibinfo {author} {\bibfnamefont
  {Y.}~\bibnamefont {Zhang}},\ }\href {\doibase 10.1038/s41586-018-0626-9}
  {\bibfield  {journal} {\bibinfo  {journal} {Nature}\ }\textbf {\bibinfo
  {volume} {563}},\ \bibinfo {pages} {94} (\bibinfo {year} {2018})}\BibitemShut
  {NoStop}%
\bibitem [{\citenamefont {Fei}\ \emph {et~al.}(2018)\citenamefont {Fei},
  \citenamefont {Huang}, \citenamefont {Malinowski}, \citenamefont {Wang},
  \citenamefont {Song}, \citenamefont {Sanchez}, \citenamefont {Yao},
  \citenamefont {Xiao}, \citenamefont {Zhu}, \citenamefont {May} \emph
  {et~al.}}]{fei2018two}%
  \BibitemOpen
  \bibfield  {author} {\bibinfo {author} {\bibfnamefont {Z.}~\bibnamefont
  {Fei}}, \bibinfo {author} {\bibfnamefont {B.}~\bibnamefont {Huang}}, \bibinfo
  {author} {\bibfnamefont {P.}~\bibnamefont {Malinowski}}, \bibinfo {author}
  {\bibfnamefont {W.}~\bibnamefont {Wang}}, \bibinfo {author} {\bibfnamefont
  {T.}~\bibnamefont {Song}}, \bibinfo {author} {\bibfnamefont {J.}~\bibnamefont
  {Sanchez}}, \bibinfo {author} {\bibfnamefont {W.}~\bibnamefont {Yao}},
  \bibinfo {author} {\bibfnamefont {D.}~\bibnamefont {Xiao}}, \bibinfo {author}
  {\bibfnamefont {X.}~\bibnamefont {Zhu}}, \bibinfo {author} {\bibfnamefont
  {A.~F.}\ \bibnamefont {May}},  \emph {et~al.},\ }\href@noop {} {\bibfield
  {journal} {\bibinfo  {journal} {Nat. Mater.}\ }\textbf {\bibinfo {volume}
  {17}},\ \bibinfo {pages} {778} (\bibinfo {year} {2018})}\BibitemShut
  {NoStop}%
\bibitem [{\citenamefont {Kim}\ \emph {et~al.}(2018)\citenamefont {Kim},
  \citenamefont {Seo}, \citenamefont {Lee}, \citenamefont {Ko}, \citenamefont
  {Kim}, \citenamefont {Jang}, \citenamefont {Ok}, \citenamefont {Lee},
  \citenamefont {Jo}, \citenamefont {Kang}, \citenamefont {Shim}, \citenamefont
  {Kim}, \citenamefont {Yeom}, \citenamefont {{Il Min}}, \citenamefont {Yang},\
  and\ \citenamefont {Kim}}]{Kim2018}%
  \BibitemOpen
  \bibfield  {author} {\bibinfo {author} {\bibfnamefont {K.}~\bibnamefont
  {Kim}}, \bibinfo {author} {\bibfnamefont {J.}~\bibnamefont {Seo}}, \bibinfo
  {author} {\bibfnamefont {E.}~\bibnamefont {Lee}}, \bibinfo {author}
  {\bibfnamefont {K.~T.}\ \bibnamefont {Ko}}, \bibinfo {author} {\bibfnamefont
  {B.~S.}\ \bibnamefont {Kim}}, \bibinfo {author} {\bibfnamefont {B.~G.}\
  \bibnamefont {Jang}}, \bibinfo {author} {\bibfnamefont {J.~M.}\ \bibnamefont
  {Ok}}, \bibinfo {author} {\bibfnamefont {J.}~\bibnamefont {Lee}}, \bibinfo
  {author} {\bibfnamefont {Y.~J.}\ \bibnamefont {Jo}}, \bibinfo {author}
  {\bibfnamefont {W.}~\bibnamefont {Kang}}, \bibinfo {author} {\bibfnamefont
  {J.~H.}\ \bibnamefont {Shim}}, \bibinfo {author} {\bibfnamefont
  {C.}~\bibnamefont {Kim}}, \bibinfo {author} {\bibfnamefont {H.~W.}\
  \bibnamefont {Yeom}}, \bibinfo {author} {\bibfnamefont {B.}~\bibnamefont {{Il
  Min}}}, \bibinfo {author} {\bibfnamefont {B.~J.}\ \bibnamefont {Yang}}, \
  and\ \bibinfo {author} {\bibfnamefont {J.~S.}\ \bibnamefont {Kim}},\ }\href
  {\doibase 10.1038/s41563-018-0132-3} {\bibfield  {journal} {\bibinfo
  {journal} {Nat. Mater.}\ }\textbf {\bibinfo {volume} {17}},\ \bibinfo {pages}
  {794} (\bibinfo {year} {2018})}\BibitemShut {NoStop}%
\bibitem [{\citenamefont {Nguyen}\ \emph {et~al.}(2018)\citenamefont {Nguyen},
  \citenamefont {Lee}, \citenamefont {Berlijn}, \citenamefont {Zou},
  \citenamefont {Hus}, \citenamefont {Park}, \citenamefont {Gai}, \citenamefont
  {Lee},\ and\ \citenamefont {Li}}]{nguyen2018visualization}%
  \BibitemOpen
  \bibfield  {author} {\bibinfo {author} {\bibfnamefont {G.~D.}\ \bibnamefont
  {Nguyen}}, \bibinfo {author} {\bibfnamefont {J.}~\bibnamefont {Lee}},
  \bibinfo {author} {\bibfnamefont {T.}~\bibnamefont {Berlijn}}, \bibinfo
  {author} {\bibfnamefont {Q.}~\bibnamefont {Zou}}, \bibinfo {author}
  {\bibfnamefont {S.~M.}\ \bibnamefont {Hus}}, \bibinfo {author} {\bibfnamefont
  {J.}~\bibnamefont {Park}}, \bibinfo {author} {\bibfnamefont {Z.}~\bibnamefont
  {Gai}}, \bibinfo {author} {\bibfnamefont {C.}~\bibnamefont {Lee}}, \ and\
  \bibinfo {author} {\bibfnamefont {A.-P.}\ \bibnamefont {Li}},\ }\href@noop {}
  {\bibfield  {journal} {\bibinfo  {journal} {Phys. Rev. B}\ }\textbf {\bibinfo
  {volume} {97}},\ \bibinfo {pages} {014425} (\bibinfo {year}
  {2018})}\BibitemShut {NoStop}%
\bibitem [{\citenamefont {Ding}\ \emph {et~al.}(2019)\citenamefont {Ding},
  \citenamefont {Li}, \citenamefont {Xu}, \citenamefont {Li}, \citenamefont
  {Hou}, \citenamefont {Liu}, \citenamefont {Xi}, \citenamefont {Xu},
  \citenamefont {Yao},\ and\ \citenamefont {Wang}}]{ding2019observation}%
  \BibitemOpen
  \bibfield  {author} {\bibinfo {author} {\bibfnamefont {B.}~\bibnamefont
  {Ding}}, \bibinfo {author} {\bibfnamefont {Z.}~\bibnamefont {Li}}, \bibinfo
  {author} {\bibfnamefont {G.}~\bibnamefont {Xu}}, \bibinfo {author}
  {\bibfnamefont {H.}~\bibnamefont {Li}}, \bibinfo {author} {\bibfnamefont
  {Z.}~\bibnamefont {Hou}}, \bibinfo {author} {\bibfnamefont {E.}~\bibnamefont
  {Liu}}, \bibinfo {author} {\bibfnamefont {X.}~\bibnamefont {Xi}}, \bibinfo
  {author} {\bibfnamefont {F.}~\bibnamefont {Xu}}, \bibinfo {author}
  {\bibfnamefont {Y.}~\bibnamefont {Yao}}, \ and\ \bibinfo {author}
  {\bibfnamefont {W.}~\bibnamefont {Wang}},\ }\href@noop {} {\bibfield
  {journal} {\bibinfo  {journal} {Nano Lett.}\ }\textbf {\bibinfo {volume}
  {20}},\ \bibinfo {pages} {868} (\bibinfo {year} {2019})}\BibitemShut
  {NoStop}%
\bibitem [{\citenamefont {Shabbir}\ \emph {et~al.}(2018)\citenamefont
  {Shabbir}, \citenamefont {Nadeem}, \citenamefont {Dai}, \citenamefont
  {Fuhrer}, \citenamefont {Xue}, \citenamefont {Wang},\ and\ \citenamefont
  {Bao}}]{shabbir2018long}%
  \BibitemOpen
  \bibfield  {author} {\bibinfo {author} {\bibfnamefont {B.}~\bibnamefont
  {Shabbir}}, \bibinfo {author} {\bibfnamefont {M.}~\bibnamefont {Nadeem}},
  \bibinfo {author} {\bibfnamefont {Z.}~\bibnamefont {Dai}}, \bibinfo {author}
  {\bibfnamefont {M.~S.}\ \bibnamefont {Fuhrer}}, \bibinfo {author}
  {\bibfnamefont {Q.-K.}\ \bibnamefont {Xue}}, \bibinfo {author} {\bibfnamefont
  {X.}~\bibnamefont {Wang}}, \ and\ \bibinfo {author} {\bibfnamefont
  {Q.}~\bibnamefont {Bao}},\ }\href@noop {} {\bibfield  {journal} {\bibinfo
  {journal} {Appl. Phys. Rev.}\ }\textbf {\bibinfo {volume} {5}},\ \bibinfo
  {pages} {041105} (\bibinfo {year} {2018})}\BibitemShut {NoStop}%
\bibitem [{\citenamefont {Le{\'o}n-Brito}\ \emph {et~al.}(2016)\citenamefont
  {Le{\'o}n-Brito}, \citenamefont {Bauer}, \citenamefont {Ronning},
  \citenamefont {Thompson},\ and\ \citenamefont
  {Movshovich}}]{leon2016magnetic}%
  \BibitemOpen
  \bibfield  {author} {\bibinfo {author} {\bibfnamefont {N.}~\bibnamefont
  {Le{\'o}n-Brito}}, \bibinfo {author} {\bibfnamefont {E.~D.}\ \bibnamefont
  {Bauer}}, \bibinfo {author} {\bibfnamefont {F.}~\bibnamefont {Ronning}},
  \bibinfo {author} {\bibfnamefont {J.~D.}\ \bibnamefont {Thompson}}, \ and\
  \bibinfo {author} {\bibfnamefont {R.}~\bibnamefont {Movshovich}},\
  }\href@noop {} {\bibfield  {journal} {\bibinfo  {journal} {J. Appl. Phys.}\
  }\textbf {\bibinfo {volume} {120}},\ \bibinfo {pages} {083903} (\bibinfo
  {year} {2016})}\BibitemShut {NoStop}%
\bibitem [{\citenamefont {Costa}\ \emph {et~al.}(2020)\citenamefont {Costa},
  \citenamefont {Peres}, \citenamefont {Fern{\'a}ndez-Rossier},\ and\
  \citenamefont {Costa}}]{costa2020nonreciprocal}%
  \BibitemOpen
  \bibfield  {author} {\bibinfo {author} {\bibfnamefont {M.}~\bibnamefont
  {Costa}}, \bibinfo {author} {\bibfnamefont {N.~M.}\ \bibnamefont {Peres}},
  \bibinfo {author} {\bibfnamefont {J.}~\bibnamefont {Fern{\'a}ndez-Rossier}},
  \ and\ \bibinfo {author} {\bibfnamefont {A.~T.}\ \bibnamefont {Costa}},\
  }\href@noop {} {\bibfield  {journal} {\bibinfo  {journal} {Phys. Rev. B}\
  }\textbf {\bibinfo {volume} {102}},\ \bibinfo {pages} {014450} (\bibinfo
  {year} {2020})}\BibitemShut {NoStop}%
\bibitem [{\citenamefont {Hubert}\ and\ \citenamefont
  {Sch\"afer}(2000)}]{hubert-schafer2000}%
  \BibitemOpen
  \bibfield  {author} {\bibinfo {author} {\bibfnamefont {A.}~\bibnamefont
  {Hubert}}\ and\ \bibinfo {author} {\bibfnamefont {R.}~\bibnamefont
  {Sch\"afer}},\ }\href@noop {} {\emph {\bibinfo {title} {Magnetic Domains}}}\
  (\bibinfo  {publisher} {Springer},\ \bibinfo {year} {2000})\BibitemShut
  {NoStop}%
\bibitem [{\citenamefont {Deiseroth}\ \emph {et~al.}(2006)\citenamefont
  {Deiseroth}, \citenamefont {Aleksandrov}, \citenamefont {Reiner},
  \citenamefont {Kienle},\ and\ \citenamefont {Kremer}}]{Deiseroth2006}%
  \BibitemOpen
  \bibfield  {author} {\bibinfo {author} {\bibfnamefont {H.~J.}\ \bibnamefont
  {Deiseroth}}, \bibinfo {author} {\bibfnamefont {K.}~\bibnamefont
  {Aleksandrov}}, \bibinfo {author} {\bibfnamefont {C.}~\bibnamefont {Reiner}},
  \bibinfo {author} {\bibfnamefont {L.}~\bibnamefont {Kienle}}, \ and\ \bibinfo
  {author} {\bibfnamefont {R.~K.}\ \bibnamefont {Kremer}},\ }\href {\doibase
  10.1002/ejic.200501020} {\bibfield  {journal} {\bibinfo  {journal} {Eur. J.
  Inorg. Chem.}\ }\textbf {\bibinfo {volume} {2006}},\ \bibinfo {pages} {1561}
  (\bibinfo {year} {2006})}\BibitemShut {NoStop}%
\bibitem [{\citenamefont {Chen}\ \emph
  {et~al.}(2013{\natexlab{a}})\citenamefont {Chen}, \citenamefont {Yang},
  \citenamefont {Wang}, \citenamefont {Imai}, \citenamefont {Ohta},
  \citenamefont {Michioka}, \citenamefont {Yoshimura},\ and\ \citenamefont
  {Fang}}]{chen2013magnetic}%
  \BibitemOpen
  \bibfield  {author} {\bibinfo {author} {\bibfnamefont {B.}~\bibnamefont
  {Chen}}, \bibinfo {author} {\bibfnamefont {J.}~\bibnamefont {Yang}}, \bibinfo
  {author} {\bibfnamefont {H.}~\bibnamefont {Wang}}, \bibinfo {author}
  {\bibfnamefont {M.}~\bibnamefont {Imai}}, \bibinfo {author} {\bibfnamefont
  {H.}~\bibnamefont {Ohta}}, \bibinfo {author} {\bibfnamefont {C.}~\bibnamefont
  {Michioka}}, \bibinfo {author} {\bibfnamefont {K.}~\bibnamefont {Yoshimura}},
  \ and\ \bibinfo {author} {\bibfnamefont {M.}~\bibnamefont {Fang}},\
  }\href@noop {} {\bibfield  {journal} {\bibinfo  {journal} {J. Phys. Soc.
  Japan}\ }\textbf {\bibinfo {volume} {82}},\ \bibinfo {pages} {124711}
  (\bibinfo {year} {2013}{\natexlab{a}})}\BibitemShut {NoStop}%
\bibitem [{\citenamefont {Wachowiak}\ \emph {et~al.}(2002)\citenamefont
  {Wachowiak}, \citenamefont {Wiebe}, \citenamefont {Bode}, \citenamefont
  {Pietzsch}, \citenamefont {Morgenstern},\ and\ \citenamefont
  {Wiesendanger}}]{Wachowiak2002a}%
  \BibitemOpen
  \bibfield  {author} {\bibinfo {author} {\bibfnamefont {A.}~\bibnamefont
  {Wachowiak}}, \bibinfo {author} {\bibfnamefont {J.}~\bibnamefont {Wiebe}},
  \bibinfo {author} {\bibfnamefont {M.}~\bibnamefont {Bode}}, \bibinfo {author}
  {\bibfnamefont {O.}~\bibnamefont {Pietzsch}}, \bibinfo {author}
  {\bibfnamefont {M.}~\bibnamefont {Morgenstern}}, \ and\ \bibinfo {author}
  {\bibfnamefont {R.}~\bibnamefont {Wiesendanger}},\ }\href {\doibase
  10.1126/science.1075302} {\bibfield  {journal} {\bibinfo  {journal}
  {Science}\ }\textbf {\bibinfo {volume} {298}},\ \bibinfo {pages} {577}
  (\bibinfo {year} {2002})}\BibitemShut {NoStop}%
\bibitem [{\citenamefont {Julliere}(1975)}]{Julliere1975}%
  \BibitemOpen
  \bibfield  {author} {\bibinfo {author} {\bibfnamefont {M.}~\bibnamefont
  {Julliere}},\ }\href {\doibase 10.1016/0375-9601(75)90174-7} {\bibfield
  {journal} {\bibinfo  {journal} {Phys. Lett.}\ }\textbf {\bibinfo {volume}
  {54A}},\ \bibinfo {pages} {225} (\bibinfo {year} {1975})}\BibitemShut
  {NoStop}%
\bibitem [{\citenamefont {Wiesendanger}(2009)}]{Wiesendanger2009spin}%
  \BibitemOpen
  \bibfield  {author} {\bibinfo {author} {\bibfnamefont {R.}~\bibnamefont
  {Wiesendanger}},\ }\href@noop {} {\bibfield  {journal} {\bibinfo  {journal}
  {Rev. Mod. Phys.}\ }\textbf {\bibinfo {volume} {81}},\ \bibinfo {pages}
  {1495} (\bibinfo {year} {2009})}\BibitemShut {NoStop}%
\bibitem [{\citenamefont {Wulfhekel}\ and\ \citenamefont
  {Kirschner}(2007)}]{Wulfhekel2007}%
  \BibitemOpen
  \bibfield  {author} {\bibinfo {author} {\bibfnamefont {W.}~\bibnamefont
  {Wulfhekel}}\ and\ \bibinfo {author} {\bibfnamefont {J.}~\bibnamefont
  {Kirschner}},\ }\href {\doibase 10.1146/ANNUREV.MATSCI.37.052506.084342}
  {\bibfield  {journal} {\bibinfo  {journal} {Annu. Rev. Mater. Res.}\ }\textbf
  {\bibinfo {volume} {37}},\ \bibinfo {pages} {69} (\bibinfo {year}
  {2007})}\BibitemShut {NoStop}%
\bibitem [{\citenamefont {Menzel}\ \emph {et~al.}(2012)\citenamefont {Menzel},
  \citenamefont {Mokrousov}, \citenamefont {Wieser}, \citenamefont {Jessica},
  \citenamefont {Vedmedenko}, \citenamefont {Bl}, \citenamefont {Heinze},
  \citenamefont {Bickel}, \citenamefont {Vedmedenko}, \citenamefont {Blu},
  \citenamefont {Wiesendanger}, \citenamefont {Heinze},\ and\ \citenamefont
  {Bergmann}}]{Menzel2012}%
  \BibitemOpen
  \bibfield  {author} {\bibinfo {author} {\bibfnamefont {M.}~\bibnamefont
  {Menzel}}, \bibinfo {author} {\bibfnamefont {Y.}~\bibnamefont {Mokrousov}},
  \bibinfo {author} {\bibfnamefont {R.}~\bibnamefont {Wieser}}, \bibinfo
  {author} {\bibfnamefont {E.}~\bibnamefont {Jessica}}, \bibinfo {author}
  {\bibfnamefont {E.}~\bibnamefont {Vedmedenko}}, \bibinfo {author}
  {\bibfnamefont {S.}~\bibnamefont {Bl}}, \bibinfo {author} {\bibfnamefont
  {S.}~\bibnamefont {Heinze}}, \bibinfo {author} {\bibfnamefont {J.~E.}\
  \bibnamefont {Bickel}}, \bibinfo {author} {\bibfnamefont {E.}~\bibnamefont
  {Vedmedenko}}, \bibinfo {author} {\bibfnamefont {S.}~\bibnamefont {Blu}},
  \bibinfo {author} {\bibfnamefont {R.}~\bibnamefont {Wiesendanger}}, \bibinfo
  {author} {\bibfnamefont {S.}~\bibnamefont {Heinze}}, \ and\ \bibinfo {author}
  {\bibfnamefont {K.~V.}\ \bibnamefont {Bergmann}},\ }\href {\doibase
  10.1103/PhysRevLett.108.197204} {\bibfield  {journal} {\bibinfo  {journal}
  {Phys. Rev. Lett.}\ }\textbf {\bibinfo {volume} {108}},\ \bibinfo {pages}
  {197204} (\bibinfo {year} {2012})}\BibitemShut {NoStop}%
\bibitem [{\citenamefont {Tersoff}\ and\ \citenamefont
  {Hamann}(1985)}]{Tersoff1985}%
  \BibitemOpen
  \bibfield  {author} {\bibinfo {author} {\bibfnamefont {J.}~\bibnamefont
  {Tersoff}}\ and\ \bibinfo {author} {\bibfnamefont {D.~R.}\ \bibnamefont
  {Hamann}},\ }\href {\doibase 10.1103/PhysRevB.31.805} {\bibfield  {journal}
  {\bibinfo  {journal} {Phys. Rev. B}\ }\textbf {\bibinfo {volume} {31}},\
  \bibinfo {pages} {805} (\bibinfo {year} {1985})}\BibitemShut {NoStop}%
\bibitem [{\citenamefont {Moodera}\ \emph {et~al.}(1998)\citenamefont
  {Moodera}, \citenamefont {Nowak},\ and\ \citenamefont {van~de
  Veerdonk}}]{Moodera1998}%
  \BibitemOpen
  \bibfield  {author} {\bibinfo {author} {\bibfnamefont {J.~S.}\ \bibnamefont
  {Moodera}}, \bibinfo {author} {\bibfnamefont {J.}~\bibnamefont {Nowak}}, \
  and\ \bibinfo {author} {\bibfnamefont {R.~J.~M.}\ \bibnamefont {van~de
  Veerdonk}},\ }\href {\doibase 10.1103/PhysRevLett.80.2941} {\bibfield
  {journal} {\bibinfo  {journal} {Phys. Rev. Lett.}\ }\textbf {\bibinfo
  {volume} {80}},\ \bibinfo {pages} {2941} (\bibinfo {year}
  {1998})}\BibitemShut {NoStop}%
\bibitem [{\citenamefont {Balashov}\ \emph {et~al.}(2006)\citenamefont
  {Balashov}, \citenamefont {Tak{\'{a}}cs}, \citenamefont {Wulfhekel},\ and\
  \citenamefont {Kirschner}}]{Balashov2006}%
  \BibitemOpen
  \bibfield  {author} {\bibinfo {author} {\bibfnamefont {T.}~\bibnamefont
  {Balashov}}, \bibinfo {author} {\bibfnamefont {A.~F.}\ \bibnamefont
  {Tak{\'{a}}cs}}, \bibinfo {author} {\bibfnamefont {W.}~\bibnamefont
  {Wulfhekel}}, \ and\ \bibinfo {author} {\bibfnamefont {J.}~\bibnamefont
  {Kirschner}},\ }\href {\doibase 10.1103/PhysRevLett.97.187201} {\bibfield
  {journal} {\bibinfo  {journal} {Phys. Rev. Lett.}\ }\textbf {\bibinfo
  {volume} {97}},\ \bibinfo {pages} {187201} (\bibinfo {year}
  {2006})}\BibitemShut {NoStop}%
\bibitem [{\citenamefont {Gao}\ \emph {et~al.}(2008)\citenamefont {Gao},
  \citenamefont {Ernst}, \citenamefont {Fischer}, \citenamefont {Hergert},
  \citenamefont {Bruno}, \citenamefont {Wulfhekel},\ and\ \citenamefont
  {Kirschner}}]{Gao2008}%
  \BibitemOpen
  \bibfield  {author} {\bibinfo {author} {\bibfnamefont {C.~L.}\ \bibnamefont
  {Gao}}, \bibinfo {author} {\bibfnamefont {A.}~\bibnamefont {Ernst}}, \bibinfo
  {author} {\bibfnamefont {G.}~\bibnamefont {Fischer}}, \bibinfo {author}
  {\bibfnamefont {W.}~\bibnamefont {Hergert}}, \bibinfo {author} {\bibfnamefont
  {P.}~\bibnamefont {Bruno}}, \bibinfo {author} {\bibfnamefont
  {W.}~\bibnamefont {Wulfhekel}}, \ and\ \bibinfo {author} {\bibfnamefont
  {J.}~\bibnamefont {Kirschner}},\ }\href {\doibase
  10.1103/PhysRevLett.101.167201} {\bibfield  {journal} {\bibinfo  {journal}
  {Phys. Rev. Lett.}\ }\textbf {\bibinfo {volume} {101}},\ \bibinfo {pages}
  {167201} (\bibinfo {year} {2008})}\BibitemShut {NoStop}%
\bibitem [{\citenamefont {Balashov}\ \emph {et~al.}(2008)\citenamefont
  {Balashov}, \citenamefont {Tak{\'{a}}cs}, \citenamefont {D{\"{a}}ne},
  \citenamefont {Ernst}, \citenamefont {Bruno},\ and\ \citenamefont
  {Wulfhekel}}]{Balashov2008}%
  \BibitemOpen
  \bibfield  {author} {\bibinfo {author} {\bibfnamefont {T.}~\bibnamefont
  {Balashov}}, \bibinfo {author} {\bibfnamefont {A.~F.}\ \bibnamefont
  {Tak{\'{a}}cs}}, \bibinfo {author} {\bibfnamefont {M.}~\bibnamefont
  {D{\"{a}}ne}}, \bibinfo {author} {\bibfnamefont {A.}~\bibnamefont {Ernst}},
  \bibinfo {author} {\bibfnamefont {P.}~\bibnamefont {Bruno}}, \ and\ \bibinfo
  {author} {\bibfnamefont {W.}~\bibnamefont {Wulfhekel}},\ }\href {\doibase
  10.1103/PhysRevB.78.174404} {\bibfield  {journal} {\bibinfo  {journal} {Phys.
  Rev. B}\ }\textbf {\bibinfo {volume} {78}},\ \bibinfo {pages} {174404}
  (\bibinfo {year} {2008})}\BibitemShut {NoStop}%
\bibitem [{\citenamefont {Zhu}\ \emph {et~al.}(2021)\citenamefont {Zhu},
  \citenamefont {You}, \citenamefont {Xu}, \citenamefont {Tang}, \citenamefont
  {Gong},\ and\ \citenamefont {Xu}}]{ZHU2021107085}%
  \BibitemOpen
  \bibfield  {author} {\bibinfo {author} {\bibfnamefont {M.}~\bibnamefont
  {Zhu}}, \bibinfo {author} {\bibfnamefont {Y.}~\bibnamefont {You}}, \bibinfo
  {author} {\bibfnamefont {G.}~\bibnamefont {Xu}}, \bibinfo {author}
  {\bibfnamefont {J.}~\bibnamefont {Tang}}, \bibinfo {author} {\bibfnamefont
  {Y.}~\bibnamefont {Gong}}, \ and\ \bibinfo {author} {\bibfnamefont
  {F.}~\bibnamefont {Xu}},\ }\href {\doibase
  https://doi.org/10.1016/j.intermet.2021.107085} {\bibfield  {journal}
  {\bibinfo  {journal} {Intermetallics}\ }\textbf {\bibinfo {volume} {131}},\
  \bibinfo {pages} {107085} (\bibinfo {year} {2021})}\BibitemShut {NoStop}%
\bibitem [{\citenamefont {Chen}\ \emph
  {et~al.}(2013{\natexlab{b}})\citenamefont {Chen}, \citenamefont {Yang},
  \citenamefont {Wang}, \citenamefont {Imai}, \citenamefont {Ohta},
  \citenamefont {Michioka}, \citenamefont {Yoshimura},\ and\ \citenamefont
  {Fang}}]{doi:10.7566/JPSJ.82.124711}%
  \BibitemOpen
  \bibfield  {author} {\bibinfo {author} {\bibfnamefont {B.}~\bibnamefont
  {Chen}}, \bibinfo {author} {\bibfnamefont {J.}~\bibnamefont {Yang}}, \bibinfo
  {author} {\bibfnamefont {H.}~\bibnamefont {Wang}}, \bibinfo {author}
  {\bibfnamefont {M.}~\bibnamefont {Imai}}, \bibinfo {author} {\bibfnamefont
  {H.}~\bibnamefont {Ohta}}, \bibinfo {author} {\bibfnamefont {C.}~\bibnamefont
  {Michioka}}, \bibinfo {author} {\bibfnamefont {K.}~\bibnamefont {Yoshimura}},
  \ and\ \bibinfo {author} {\bibfnamefont {M.}~\bibnamefont {Fang}},\ }\href
  {\doibase 10.7566/JPSJ.82.124711} {\bibfield  {journal} {\bibinfo  {journal}
  {J. Phys. Soc. Japan.}\ }\textbf {\bibinfo {volume} {82}},\ \bibinfo {pages}
  {124711} (\bibinfo {year} {2013}{\natexlab{b}})}\BibitemShut {NoStop}%
\bibitem [{\citenamefont {May}\ \emph {et~al.}(2016)\citenamefont {May},
  \citenamefont {Calder}, \citenamefont {Cantoni}, \citenamefont {Cao},\ and\
  \citenamefont {McGuire}}]{May2016}%
  \BibitemOpen
  \bibfield  {author} {\bibinfo {author} {\bibfnamefont {A.~F.}\ \bibnamefont
  {May}}, \bibinfo {author} {\bibfnamefont {S.}~\bibnamefont {Calder}},
  \bibinfo {author} {\bibfnamefont {C.}~\bibnamefont {Cantoni}}, \bibinfo
  {author} {\bibfnamefont {H.}~\bibnamefont {Cao}}, \ and\ \bibinfo {author}
  {\bibfnamefont {M.~A.}\ \bibnamefont {McGuire}},\ }\href {\doibase
  10.1103/PhysRevB.93.014411} {\bibfield  {journal} {\bibinfo  {journal} {Phys.
  Rev. B}\ }\textbf {\bibinfo {volume} {93}},\ \bibinfo {pages} {014411}
  (\bibinfo {year} {2016})}\BibitemShut {NoStop}%
\bibitem [{\citenamefont {Scheinfein}\ \emph {et~al.}(1991)\citenamefont
  {Scheinfein}, \citenamefont {Unguris}, \citenamefont {Blue}, \citenamefont
  {Coakley}, \citenamefont {Pierce}, \citenamefont {Celotta},\ and\
  \citenamefont {Ryan}}]{Scheinfein1991}%
  \BibitemOpen
  \bibfield  {author} {\bibinfo {author} {\bibfnamefont {M.~R.}\ \bibnamefont
  {Scheinfein}}, \bibinfo {author} {\bibfnamefont {J.}~\bibnamefont {Unguris}},
  \bibinfo {author} {\bibfnamefont {J.~L.}\ \bibnamefont {Blue}}, \bibinfo
  {author} {\bibfnamefont {K.~J.}\ \bibnamefont {Coakley}}, \bibinfo {author}
  {\bibfnamefont {D.~T.}\ \bibnamefont {Pierce}}, \bibinfo {author}
  {\bibfnamefont {R.~J.}\ \bibnamefont {Celotta}}, \ and\ \bibinfo {author}
  {\bibfnamefont {P.~J.}\ \bibnamefont {Ryan}},\ }\href {\doibase
  10.1103/PhysRevB.43.3395} {\bibfield  {journal} {\bibinfo  {journal} {Phys.
  Rev. B}\ }\textbf {\bibinfo {volume} {43}},\ \bibinfo {pages} {3395}
  (\bibinfo {year} {1991})}\BibitemShut {NoStop}%
\bibitem [{\citenamefont {Schlickum}\ \emph {et~al.}(2003)\citenamefont
  {Schlickum}, \citenamefont {Wulfhekel},\ and\ \citenamefont
  {Kirschner}}]{Schlickum2003}%
  \BibitemOpen
  \bibfield  {author} {\bibinfo {author} {\bibfnamefont {U.}~\bibnamefont
  {Schlickum}}, \bibinfo {author} {\bibfnamefont {W.}~\bibnamefont
  {Wulfhekel}}, \ and\ \bibinfo {author} {\bibfnamefont {J.}~\bibnamefont
  {Kirschner}},\ }\href {\doibase 10.1063/1.1606867} {\bibfield  {journal}
  {\bibinfo  {journal} {Appl. Phys. Lett.}\ }\textbf {\bibinfo {volume} {83}},\
  \bibinfo {pages} {2016} (\bibinfo {year} {2003})}\BibitemShut {NoStop}%
\bibitem [{\citenamefont {Schweflinghaus}\ \emph {et~al.}(2016)\citenamefont
  {Schweflinghaus}, \citenamefont {Zimmermann}, \citenamefont {Heide},
  \citenamefont {Bihlmayer},\ and\ \citenamefont
  {Bl\"ugel}}]{Schweflinghaus2016}%
  \BibitemOpen
  \bibfield  {author} {\bibinfo {author} {\bibfnamefont {B.}~\bibnamefont
  {Schweflinghaus}}, \bibinfo {author} {\bibfnamefont {B.}~\bibnamefont
  {Zimmermann}}, \bibinfo {author} {\bibfnamefont {M.}~\bibnamefont {Heide}},
  \bibinfo {author} {\bibfnamefont {G.}~\bibnamefont {Bihlmayer}}, \ and\
  \bibinfo {author} {\bibfnamefont {S.}~\bibnamefont {Bl\"ugel}},\ }\href
  {\doibase 10.1103/PhysRevB.94.024403} {\bibfield  {journal} {\bibinfo
  {journal} {Phys. Rev. B}\ }\textbf {\bibinfo {volume} {94}},\ \bibinfo
  {pages} {024403} (\bibinfo {year} {2016})}\BibitemShut {NoStop}%
\bibitem [{\citenamefont {Smolenski{\u\i}}\ and\ \citenamefont
  {Chupis}(1982)}]{smolenskiui1982ferroelectromagnets}%
  \BibitemOpen
  \bibfield  {author} {\bibinfo {author} {\bibfnamefont {G.~A.}\ \bibnamefont
  {Smolenski{\u\i}}}\ and\ \bibinfo {author} {\bibfnamefont {I.~E.}\
  \bibnamefont {Chupis}},\ }\href@noop {} {\bibfield  {journal} {\bibinfo
  {journal} {Sov. Phys. Uspekhi}\ }\textbf {\bibinfo {volume} {25}},\ \bibinfo
  {pages} {475} (\bibinfo {year} {1982})}\BibitemShut {NoStop}%
\bibitem [{\citenamefont {Lottermoser}\ \emph {et~al.}(2004)\citenamefont
  {Lottermoser}, \citenamefont {Lonkai}, \citenamefont {Amann}, \citenamefont
  {Hohlwein}, \citenamefont {Ihringer},\ and\ \citenamefont
  {Fiebig}}]{lottermoser2004magnetic}%
  \BibitemOpen
  \bibfield  {author} {\bibinfo {author} {\bibfnamefont {T.}~\bibnamefont
  {Lottermoser}}, \bibinfo {author} {\bibfnamefont {T.}~\bibnamefont {Lonkai}},
  \bibinfo {author} {\bibfnamefont {U.}~\bibnamefont {Amann}}, \bibinfo
  {author} {\bibfnamefont {D.}~\bibnamefont {Hohlwein}}, \bibinfo {author}
  {\bibfnamefont {J.}~\bibnamefont {Ihringer}}, \ and\ \bibinfo {author}
  {\bibfnamefont {M.}~\bibnamefont {Fiebig}},\ }\href@noop {} {\bibfield
  {journal} {\bibinfo  {journal} {Nature}\ }\textbf {\bibinfo {volume} {430}},\
  \bibinfo {pages} {541} (\bibinfo {year} {2004})}\BibitemShut {NoStop}%
\bibitem [{\citenamefont {Spaldin}\ and\ \citenamefont
  {Fiebig}(2005)}]{spaldin2005renaissance}%
  \BibitemOpen
  \bibfield  {author} {\bibinfo {author} {\bibfnamefont {N.~A.}\ \bibnamefont
  {Spaldin}}\ and\ \bibinfo {author} {\bibfnamefont {M.}~\bibnamefont
  {Fiebig}},\ }\href@noop {} {\bibfield  {journal} {\bibinfo  {journal}
  {Science}\ }\textbf {\bibinfo {volume} {309}},\ \bibinfo {pages} {391}
  (\bibinfo {year} {2005})}\BibitemShut {NoStop}%
\bibitem [{\citenamefont {Zheng}\ \emph {et~al.}(2004)\citenamefont {Zheng},
  \citenamefont {Wang}, \citenamefont {Lofland}, \citenamefont {Ma},
  \citenamefont {Mohaddes-Ardabili}, \citenamefont {Zhao}, \citenamefont
  {Salamanca-Riba}, \citenamefont {Shinde}, \citenamefont {Ogale},
  \citenamefont {Bai} \emph {et~al.}}]{zheng2004multiferroic}%
  \BibitemOpen
  \bibfield  {author} {\bibinfo {author} {\bibfnamefont {H.}~\bibnamefont
  {Zheng}}, \bibinfo {author} {\bibfnamefont {J.}~\bibnamefont {Wang}},
  \bibinfo {author} {\bibfnamefont {S.}~\bibnamefont {Lofland}}, \bibinfo
  {author} {\bibfnamefont {Z.}~\bibnamefont {Ma}}, \bibinfo {author}
  {\bibfnamefont {L.}~\bibnamefont {Mohaddes-Ardabili}}, \bibinfo {author}
  {\bibfnamefont {T.}~\bibnamefont {Zhao}}, \bibinfo {author} {\bibfnamefont
  {L.}~\bibnamefont {Salamanca-Riba}}, \bibinfo {author} {\bibfnamefont
  {S.}~\bibnamefont {Shinde}}, \bibinfo {author} {\bibfnamefont
  {S.}~\bibnamefont {Ogale}}, \bibinfo {author} {\bibfnamefont
  {F.}~\bibnamefont {Bai}},  \emph {et~al.},\ }\href@noop {} {\bibfield
  {journal} {\bibinfo  {journal} {Science}\ }\textbf {\bibinfo {volume}
  {303}},\ \bibinfo {pages} {661} (\bibinfo {year} {2004})}\BibitemShut
  {NoStop}%
\bibitem [{\citenamefont {Zavaliche}\ \emph {et~al.}(2005)\citenamefont
  {Zavaliche}, \citenamefont {Zheng}, \citenamefont {Mohaddes-Ardabili},
  \citenamefont {Yang}, \citenamefont {Zhan}, \citenamefont {Shafer},
  \citenamefont {Reilly}, \citenamefont {Chopdekar}, \citenamefont {Jia},
  \citenamefont {Wright} \emph {et~al.}}]{zavaliche2005electric}%
  \BibitemOpen
  \bibfield  {author} {\bibinfo {author} {\bibfnamefont {F.}~\bibnamefont
  {Zavaliche}}, \bibinfo {author} {\bibfnamefont {H.}~\bibnamefont {Zheng}},
  \bibinfo {author} {\bibfnamefont {L.}~\bibnamefont {Mohaddes-Ardabili}},
  \bibinfo {author} {\bibfnamefont {S.}~\bibnamefont {Yang}}, \bibinfo {author}
  {\bibfnamefont {Q.}~\bibnamefont {Zhan}}, \bibinfo {author} {\bibfnamefont
  {P.}~\bibnamefont {Shafer}}, \bibinfo {author} {\bibfnamefont
  {E.}~\bibnamefont {Reilly}}, \bibinfo {author} {\bibfnamefont
  {R.}~\bibnamefont {Chopdekar}}, \bibinfo {author} {\bibfnamefont
  {Y.}~\bibnamefont {Jia}}, \bibinfo {author} {\bibfnamefont {P.}~\bibnamefont
  {Wright}},  \emph {et~al.},\ }\href@noop {} {\bibfield  {journal} {\bibinfo
  {journal} {Nano Lett.}\ }\textbf {\bibinfo {volume} {5}},\ \bibinfo {pages}
  {1793} (\bibinfo {year} {2005})}\BibitemShut {NoStop}%
\bibitem [{\citenamefont {Duan}\ \emph {et~al.}(2006)\citenamefont {Duan},
  \citenamefont {Jaswal},\ and\ \citenamefont {Tsymbal}}]{duan2006predicted}%
  \BibitemOpen
  \bibfield  {author} {\bibinfo {author} {\bibfnamefont {C.-G.}\ \bibnamefont
  {Duan}}, \bibinfo {author} {\bibfnamefont {S.~S.}\ \bibnamefont {Jaswal}}, \
  and\ \bibinfo {author} {\bibfnamefont {E.~Y.}\ \bibnamefont {Tsymbal}},\
  }\href@noop {} {\bibfield  {journal} {\bibinfo  {journal} {Phys. Rev. Lett.}\
  }\textbf {\bibinfo {volume} {97}},\ \bibinfo {pages} {047201} (\bibinfo
  {year} {2006})}\BibitemShut {NoStop}%
\bibitem [{\citenamefont {Weisheit}\ \emph {et~al.}(2007)\citenamefont
  {Weisheit}, \citenamefont {F{\"{a}}hler}, \citenamefont {Marty},
  \citenamefont {Souche}, \citenamefont {Poinsignon},\ and\ \citenamefont
  {Givord}}]{Weisheit2007}%
  \BibitemOpen
  \bibfield  {author} {\bibinfo {author} {\bibfnamefont {M.}~\bibnamefont
  {Weisheit}}, \bibinfo {author} {\bibfnamefont {S.}~\bibnamefont
  {F{\"{a}}hler}}, \bibinfo {author} {\bibfnamefont {A.}~\bibnamefont {Marty}},
  \bibinfo {author} {\bibfnamefont {Y.}~\bibnamefont {Souche}}, \bibinfo
  {author} {\bibfnamefont {C.}~\bibnamefont {Poinsignon}}, \ and\ \bibinfo
  {author} {\bibfnamefont {D.}~\bibnamefont {Givord}},\ }\href {\doibase
  10.1126/SCIENCE.1136629/SUPPL_FILE/WEISHEIT.SOM.PDF} {\bibfield  {journal}
  {\bibinfo  {journal} {Science}\ }\textbf {\bibinfo {volume} {315}},\ \bibinfo
  {pages} {349} (\bibinfo {year} {2007})}\BibitemShut {NoStop}%
\bibitem [{\citenamefont {Duan}\ \emph {et~al.}(2008)\citenamefont {Duan},
  \citenamefont {Velev}, \citenamefont {Sabirianov}, \citenamefont {Zhu},
  \citenamefont {Chu}, \citenamefont {Jaswal},\ and\ \citenamefont
  {Tsymbal}}]{Duan2008}%
  \BibitemOpen
  \bibfield  {author} {\bibinfo {author} {\bibfnamefont {C.~G.}\ \bibnamefont
  {Duan}}, \bibinfo {author} {\bibfnamefont {J.~P.}\ \bibnamefont {Velev}},
  \bibinfo {author} {\bibfnamefont {R.~F.}\ \bibnamefont {Sabirianov}},
  \bibinfo {author} {\bibfnamefont {Z.}~\bibnamefont {Zhu}}, \bibinfo {author}
  {\bibfnamefont {J.}~\bibnamefont {Chu}}, \bibinfo {author} {\bibfnamefont
  {S.~S.}\ \bibnamefont {Jaswal}}, \ and\ \bibinfo {author} {\bibfnamefont
  {E.~Y.}\ \bibnamefont {Tsymbal}},\ }\href {\doibase
  10.1103/PHYSREVLETT.101.137201/FIGURES/4/MEDIUM} {\bibfield  {journal}
  {\bibinfo  {journal} {Phys. Rev. Lett.}\ }\textbf {\bibinfo {volume} {101}},\
  \bibinfo {pages} {137201} (\bibinfo {year} {2008})}\BibitemShut {NoStop}%
\bibitem [{\citenamefont {Fechner}\ \emph {et~al.}(2008)\citenamefont
  {Fechner}, \citenamefont {Maznichenko}, \citenamefont {Ostanin},
  \citenamefont {Ernst}, \citenamefont {Henk}, \citenamefont {Bruno},\ and\
  \citenamefont {Mertig}}]{fechner2008magnetic}%
  \BibitemOpen
  \bibfield  {author} {\bibinfo {author} {\bibfnamefont {M.}~\bibnamefont
  {Fechner}}, \bibinfo {author} {\bibfnamefont {I.}~\bibnamefont
  {Maznichenko}}, \bibinfo {author} {\bibfnamefont {S.}~\bibnamefont
  {Ostanin}}, \bibinfo {author} {\bibfnamefont {A.}~\bibnamefont {Ernst}},
  \bibinfo {author} {\bibfnamefont {J.}~\bibnamefont {Henk}}, \bibinfo {author}
  {\bibfnamefont {P.}~\bibnamefont {Bruno}}, \ and\ \bibinfo {author}
  {\bibfnamefont {I.}~\bibnamefont {Mertig}},\ }\href@noop {} {\bibfield
  {journal} {\bibinfo  {journal} {Phys. Rev. B}\ }\textbf {\bibinfo {volume}
  {78}},\ \bibinfo {pages} {212406} (\bibinfo {year} {2008})}\BibitemShut
  {NoStop}%
\bibitem [{\citenamefont {Maruyama}\ \emph {et~al.}(2009)\citenamefont
  {Maruyama}, \citenamefont {Shiota}, \citenamefont {Nozaki}, \citenamefont
  {Ohta}, \citenamefont {Toda}, \citenamefont {Mizuguchi}, \citenamefont
  {Tulapurkar}, \citenamefont {Shinjo}, \citenamefont {Shiraishi},
  \citenamefont {Mizukami} \emph {et~al.}}]{maruyama2009large}%
  \BibitemOpen
  \bibfield  {author} {\bibinfo {author} {\bibfnamefont {T.}~\bibnamefont
  {Maruyama}}, \bibinfo {author} {\bibfnamefont {Y.}~\bibnamefont {Shiota}},
  \bibinfo {author} {\bibfnamefont {T.}~\bibnamefont {Nozaki}}, \bibinfo
  {author} {\bibfnamefont {K.}~\bibnamefont {Ohta}}, \bibinfo {author}
  {\bibfnamefont {N.}~\bibnamefont {Toda}}, \bibinfo {author} {\bibfnamefont
  {M.}~\bibnamefont {Mizuguchi}}, \bibinfo {author} {\bibfnamefont
  {A.}~\bibnamefont {Tulapurkar}}, \bibinfo {author} {\bibfnamefont
  {T.}~\bibnamefont {Shinjo}}, \bibinfo {author} {\bibfnamefont
  {M.}~\bibnamefont {Shiraishi}}, \bibinfo {author} {\bibfnamefont
  {S.}~\bibnamefont {Mizukami}},  \emph {et~al.},\ }\href@noop {} {\bibfield
  {journal} {\bibinfo  {journal} {Nat. Nanotechnol.}\ }\textbf {\bibinfo
  {volume} {4}},\ \bibinfo {pages} {158} (\bibinfo {year} {2009})}\BibitemShut
  {NoStop}%
\bibitem [{\citenamefont {Subkow}\ and\ \citenamefont
  {F{\"a}hnle}(2011)}]{subkow2011electron}%
  \BibitemOpen
  \bibfield  {author} {\bibinfo {author} {\bibfnamefont {S.}~\bibnamefont
  {Subkow}}\ and\ \bibinfo {author} {\bibfnamefont {M.}~\bibnamefont
  {F{\"a}hnle}},\ }\href@noop {} {\bibfield  {journal} {\bibinfo  {journal}
  {Phys. Rev. B}\ }\textbf {\bibinfo {volume} {84}},\ \bibinfo {pages} {054443}
  (\bibinfo {year} {2011})}\BibitemShut {NoStop}%
\bibitem [{\citenamefont {Shiota}\ \emph {et~al.}(2012)\citenamefont {Shiota},
  \citenamefont {Nozaki}, \citenamefont {Bonell}, \citenamefont {Murakami},
  \citenamefont {Shinjo},\ and\ \citenamefont {Suzuki}}]{shiota2012induction}%
  \BibitemOpen
  \bibfield  {author} {\bibinfo {author} {\bibfnamefont {Y.}~\bibnamefont
  {Shiota}}, \bibinfo {author} {\bibfnamefont {T.}~\bibnamefont {Nozaki}},
  \bibinfo {author} {\bibfnamefont {F.}~\bibnamefont {Bonell}}, \bibinfo
  {author} {\bibfnamefont {S.}~\bibnamefont {Murakami}}, \bibinfo {author}
  {\bibfnamefont {T.}~\bibnamefont {Shinjo}}, \ and\ \bibinfo {author}
  {\bibfnamefont {Y.}~\bibnamefont {Suzuki}},\ }\href@noop {} {\bibfield
  {journal} {\bibinfo  {journal} {Nat. Mater.}\ }\textbf {\bibinfo {volume}
  {11}},\ \bibinfo {pages} {39} (\bibinfo {year} {2012})}\BibitemShut {NoStop}%
\bibitem [{\citenamefont {Tournerie}\ \emph {et~al.}(2012)\citenamefont
  {Tournerie}, \citenamefont {Engelhardt}, \citenamefont {Maroun},\ and\
  \citenamefont {Allongue}}]{tournerie2012influence}%
  \BibitemOpen
  \bibfield  {author} {\bibinfo {author} {\bibfnamefont {N.}~\bibnamefont
  {Tournerie}}, \bibinfo {author} {\bibfnamefont {A.}~\bibnamefont
  {Engelhardt}}, \bibinfo {author} {\bibfnamefont {F.}~\bibnamefont {Maroun}},
  \ and\ \bibinfo {author} {\bibfnamefont {P.}~\bibnamefont {Allongue}},\
  }\href@noop {} {\bibfield  {journal} {\bibinfo  {journal} {Phys. Rev. B}\
  }\textbf {\bibinfo {volume} {86}},\ \bibinfo {pages} {104434} (\bibinfo
  {year} {2012})}\BibitemShut {NoStop}%
\bibitem [{\citenamefont {Nan}\ \emph {et~al.}(2014)\citenamefont {Nan},
  \citenamefont {Zhou}, \citenamefont {Liu}, \citenamefont {Yang},
  \citenamefont {Gao}, \citenamefont {Assaf}, \citenamefont {Lin},
  \citenamefont {Velu}, \citenamefont {Wang}, \citenamefont {Luo} \emph
  {et~al.}}]{nan2014}%
  \BibitemOpen
  \bibfield  {author} {\bibinfo {author} {\bibfnamefont {T.}~\bibnamefont
  {Nan}}, \bibinfo {author} {\bibfnamefont {Z.}~\bibnamefont {Zhou}}, \bibinfo
  {author} {\bibfnamefont {M.}~\bibnamefont {Liu}}, \bibinfo {author}
  {\bibfnamefont {X.}~\bibnamefont {Yang}}, \bibinfo {author} {\bibfnamefont
  {Y.}~\bibnamefont {Gao}}, \bibinfo {author} {\bibfnamefont {B.~A.}\
  \bibnamefont {Assaf}}, \bibinfo {author} {\bibfnamefont {H.}~\bibnamefont
  {Lin}}, \bibinfo {author} {\bibfnamefont {S.}~\bibnamefont {Velu}}, \bibinfo
  {author} {\bibfnamefont {X.}~\bibnamefont {Wang}}, \bibinfo {author}
  {\bibfnamefont {H.}~\bibnamefont {Luo}},  \emph {et~al.},\ }\href@noop {}
  {\bibfield  {journal} {\bibinfo  {journal} {Sci. Rep.}\ }\textbf {\bibinfo
  {volume} {4}},\ \bibinfo {pages} {3688} (\bibinfo {year} {2014})}\BibitemShut
  {NoStop}%
\bibitem [{\citenamefont {Gerhard}\ \emph {et~al.}(2014)\citenamefont
  {Gerhard}, \citenamefont {Bonell}, \citenamefont {Wulfhekel},\ and\
  \citenamefont {Suzuki}}]{gerhard2014influence}%
  \BibitemOpen
  \bibfield  {author} {\bibinfo {author} {\bibfnamefont {L.}~\bibnamefont
  {Gerhard}}, \bibinfo {author} {\bibfnamefont {F.}~\bibnamefont {Bonell}},
  \bibinfo {author} {\bibfnamefont {W.}~\bibnamefont {Wulfhekel}}, \ and\
  \bibinfo {author} {\bibfnamefont {Y.}~\bibnamefont {Suzuki}},\ }\href@noop {}
  {\bibfield  {journal} {\bibinfo  {journal} {Appl. Phys. Lett.}\ }\textbf
  {\bibinfo {volume} {105}},\ \bibinfo {pages} {152903} (\bibinfo {year}
  {2014})}\BibitemShut {NoStop}%
\bibitem [{\citenamefont {Liu}\ \emph {et~al.}(2019)\citenamefont {Liu},
  \citenamefont {Wang}, \citenamefont {Fan}, \citenamefont {Pi}, \citenamefont
  {Ge}, \citenamefont {Zhang},\ and\ \citenamefont {Zhang}}]{Liu2019}%
  \BibitemOpen
  \bibfield  {author} {\bibinfo {author} {\bibfnamefont {W.}~\bibnamefont
  {Liu}}, \bibinfo {author} {\bibfnamefont {Y.}~\bibnamefont {Wang}}, \bibinfo
  {author} {\bibfnamefont {J.}~\bibnamefont {Fan}}, \bibinfo {author}
  {\bibfnamefont {L.}~\bibnamefont {Pi}}, \bibinfo {author} {\bibfnamefont
  {M.}~\bibnamefont {Ge}}, \bibinfo {author} {\bibfnamefont {L.}~\bibnamefont
  {Zhang}}, \ and\ \bibinfo {author} {\bibfnamefont {Y.}~\bibnamefont
  {Zhang}},\ }\href {\doibase 10.1103/PhysRevB.100.104403} {\bibfield
  {journal} {\bibinfo  {journal} {Phys. Rev. B}\ }\textbf {\bibinfo {volume}
  {100}},\ \bibinfo {pages} {104403} (\bibinfo {year} {2019})}\BibitemShut
  {NoStop}%
\bibitem [{\citenamefont {Zhao}\ \emph {et~al.}(2021)\citenamefont {Zhao},
  \citenamefont {Zhao}, \citenamefont {Xi}, \citenamefont {Xu}, \citenamefont
  {Feng}, \citenamefont {Xu}, \citenamefont {Hao}, \citenamefont {Zhou},
  \citenamefont {Zhao}, \citenamefont {Dou},\ and\ \citenamefont
  {Du}}]{Zhao2021}%
  \BibitemOpen
  \bibfield  {author} {\bibinfo {author} {\bibfnamefont {M.}~\bibnamefont
  {Zhao}}, \bibinfo {author} {\bibfnamefont {Y.}~\bibnamefont {Zhao}}, \bibinfo
  {author} {\bibfnamefont {Y.}~\bibnamefont {Xi}}, \bibinfo {author}
  {\bibfnamefont {H.}~\bibnamefont {Xu}}, \bibinfo {author} {\bibfnamefont
  {H.}~\bibnamefont {Feng}}, \bibinfo {author} {\bibfnamefont {X.}~\bibnamefont
  {Xu}}, \bibinfo {author} {\bibfnamefont {W.}~\bibnamefont {Hao}}, \bibinfo
  {author} {\bibfnamefont {S.}~\bibnamefont {Zhou}}, \bibinfo {author}
  {\bibfnamefont {J.}~\bibnamefont {Zhao}}, \bibinfo {author} {\bibfnamefont
  {S.~X.}\ \bibnamefont {Dou}}, \ and\ \bibinfo {author} {\bibfnamefont
  {Y.}~\bibnamefont {Du}},\ }\href {\doibase 10.1021/acs.nanolett.1c03123}
  {\bibfield  {journal} {\bibinfo  {journal} {Nano Lett.}\ }\textbf {\bibinfo
  {volume} {21}},\ \bibinfo {pages} {9233} (\bibinfo {year}
  {2021})}\BibitemShut {NoStop}%
\bibitem [{\citenamefont {Gerhard}\ \emph {et~al.}(2010)\citenamefont
  {Gerhard}, \citenamefont {Yamada}, \citenamefont {Balashov}, \citenamefont
  {Tak{\'{a}}cs}, \citenamefont {Wesselink}, \citenamefont {D{\"{a}}ne},
  \citenamefont {Fechner}, \citenamefont {Ostanin}, \citenamefont {Ernst},
  \citenamefont {Mertig},\ and\ \citenamefont {Wulfhekel}}]{Gerhard2010}%
  \BibitemOpen
  \bibfield  {author} {\bibinfo {author} {\bibfnamefont {L.}~\bibnamefont
  {Gerhard}}, \bibinfo {author} {\bibfnamefont {T.~K.}\ \bibnamefont {Yamada}},
  \bibinfo {author} {\bibfnamefont {T.}~\bibnamefont {Balashov}}, \bibinfo
  {author} {\bibfnamefont {a.~F.}\ \bibnamefont {Tak{\'{a}}cs}}, \bibinfo
  {author} {\bibfnamefont {R.~J.~H.}\ \bibnamefont {Wesselink}}, \bibinfo
  {author} {\bibfnamefont {M.}~\bibnamefont {D{\"{a}}ne}}, \bibinfo {author}
  {\bibfnamefont {M.}~\bibnamefont {Fechner}}, \bibinfo {author} {\bibfnamefont
  {S.}~\bibnamefont {Ostanin}}, \bibinfo {author} {\bibfnamefont
  {a.}~\bibnamefont {Ernst}}, \bibinfo {author} {\bibfnamefont
  {I.}~\bibnamefont {Mertig}}, \ and\ \bibinfo {author} {\bibfnamefont
  {W.}~\bibnamefont {Wulfhekel}},\ }\href {\doibase 10.1038/nnano.2010.214}
  {\bibfield  {journal} {\bibinfo  {journal} {Nat. Nanotechnol.}\ }\textbf
  {\bibinfo {volume} {5}},\ \bibinfo {pages} {792} (\bibinfo {year}
  {2010})}\BibitemShut {NoStop}%
\bibitem [{\citenamefont {Han}\ \emph {et~al.}(2019)\citenamefont {Han},
  \citenamefont {Garlow}, \citenamefont {Liu}, \citenamefont {Zhang},
  \citenamefont {Li}, \citenamefont {DiMarzio}, \citenamefont {Knight},
  \citenamefont {Petrovic}, \citenamefont {Jariwala},\ and\ \citenamefont
  {Zhu}}]{han2019topological}%
  \BibitemOpen
  \bibfield  {author} {\bibinfo {author} {\bibfnamefont {M.-G.}\ \bibnamefont
  {Han}}, \bibinfo {author} {\bibfnamefont {J.~A.}\ \bibnamefont {Garlow}},
  \bibinfo {author} {\bibfnamefont {Y.}~\bibnamefont {Liu}}, \bibinfo {author}
  {\bibfnamefont {H.}~\bibnamefont {Zhang}}, \bibinfo {author} {\bibfnamefont
  {J.}~\bibnamefont {Li}}, \bibinfo {author} {\bibfnamefont {D.}~\bibnamefont
  {DiMarzio}}, \bibinfo {author} {\bibfnamefont {M.~W.}\ \bibnamefont
  {Knight}}, \bibinfo {author} {\bibfnamefont {C.}~\bibnamefont {Petrovic}},
  \bibinfo {author} {\bibfnamefont {D.}~\bibnamefont {Jariwala}}, \ and\
  \bibinfo {author} {\bibfnamefont {Y.}~\bibnamefont {Zhu}},\ }\href@noop {}
  {\bibfield  {journal} {\bibinfo  {journal} {Nano Lett.}\ }\textbf {\bibinfo
  {volume} {19}},\ \bibinfo {pages} {7859} (\bibinfo {year}
  {2019})}\BibitemShut {NoStop}%
\bibitem [{\citenamefont {Romming}\ \emph {et~al.}(2015)\citenamefont
  {Romming}, \citenamefont {Kubetzka}, \citenamefont {Hanneken}, \citenamefont
  {von Bergmann},\ and\ \citenamefont {Wiesendanger}}]{Romming2015}%
  \BibitemOpen
  \bibfield  {author} {\bibinfo {author} {\bibfnamefont {N.}~\bibnamefont
  {Romming}}, \bibinfo {author} {\bibfnamefont {A.}~\bibnamefont {Kubetzka}},
  \bibinfo {author} {\bibfnamefont {C.}~\bibnamefont {Hanneken}}, \bibinfo
  {author} {\bibfnamefont {K.}~\bibnamefont {von Bergmann}}, \ and\ \bibinfo
  {author} {\bibfnamefont {R.}~\bibnamefont {Wiesendanger}},\ }\href@noop {}
  {\bibfield  {journal} {\bibinfo  {journal} {Phys. Rev. Lett.}\ }\textbf
  {\bibinfo {volume} {114}},\ \bibinfo {pages} {177203} (\bibinfo {year}
  {2015})}\BibitemShut {NoStop}%
\bibitem [{\citenamefont {Rohart}\ \emph {et~al.}(2016)\citenamefont {Rohart},
  \citenamefont {Miltat},\ and\ \citenamefont {Thiaville}}]{Rohart2016}%
  \BibitemOpen
  \bibfield  {author} {\bibinfo {author} {\bibfnamefont {S.}~\bibnamefont
  {Rohart}}, \bibinfo {author} {\bibfnamefont {J.}~\bibnamefont {Miltat}}, \
  and\ \bibinfo {author} {\bibfnamefont {A.}~\bibnamefont {Thiaville}},\
  }\href@noop {} {\bibfield  {journal} {\bibinfo  {journal} {Phys. Rev. B}\
  }\textbf {\bibinfo {volume} {93}},\ \bibinfo {pages} {214412} (\bibinfo
  {year} {2016})}\BibitemShut {NoStop}%
\bibitem [{\citenamefont {Bernand-Mantel}\ \emph {et~al.}(2018)\citenamefont
  {Bernand-Mantel}, \citenamefont {Camosi}, \citenamefont {Wartelle},
  \citenamefont {Rougemaille}, \citenamefont {Darques},\ and\ \citenamefont
  {Ranno}}]{bernand2018skyrmion}%
  \BibitemOpen
  \bibfield  {author} {\bibinfo {author} {\bibfnamefont {A.}~\bibnamefont
  {Bernand-Mantel}}, \bibinfo {author} {\bibfnamefont {L.}~\bibnamefont
  {Camosi}}, \bibinfo {author} {\bibfnamefont {A.}~\bibnamefont {Wartelle}},
  \bibinfo {author} {\bibfnamefont {N.}~\bibnamefont {Rougemaille}}, \bibinfo
  {author} {\bibfnamefont {M.}~\bibnamefont {Darques}}, \ and\ \bibinfo
  {author} {\bibfnamefont {L.}~\bibnamefont {Ranno}},\ }\href@noop {}
  {\bibfield  {journal} {\bibinfo  {journal} {SciPost Phys.}\ }\textbf
  {\bibinfo {volume} {4}},\ \bibinfo {pages} {027} (\bibinfo {year}
  {2018})}\BibitemShut {NoStop}%
\bibitem [{\citenamefont {Leonov}\ \emph {et~al.}(2016)\citenamefont {Leonov},
  \citenamefont {Monchesky}, \citenamefont {Romming}, \citenamefont {Kubetzka},
  \citenamefont {Bogdanov},\ and\ \citenamefont {Wiesendanger}}]{Leonov2015a}%
  \BibitemOpen
  \bibfield  {author} {\bibinfo {author} {\bibfnamefont {A.}~\bibnamefont
  {Leonov}}, \bibinfo {author} {\bibfnamefont {T.~L.}\ \bibnamefont
  {Monchesky}}, \bibinfo {author} {\bibfnamefont {N.}~\bibnamefont {Romming}},
  \bibinfo {author} {\bibfnamefont {A.}~\bibnamefont {Kubetzka}}, \bibinfo
  {author} {\bibfnamefont {A.~N.}\ \bibnamefont {Bogdanov}}, \ and\ \bibinfo
  {author} {\bibfnamefont {R.}~\bibnamefont {Wiesendanger}},\ }\href@noop {}
  {\bibfield  {journal} {\bibinfo  {journal} {New J. Phys.}\ }\textbf {\bibinfo
  {volume} {18}},\ \bibinfo {pages} {065003} (\bibinfo {year}
  {2016})}\BibitemShut {NoStop}%
\bibitem [{\citenamefont {Mougel}\ \emph {et~al.}(2020)\citenamefont {Mougel},
  \citenamefont {Buhl}, \citenamefont {Nemoto}, \citenamefont {Balashov},
  \citenamefont {Herv{\'e}}, \citenamefont {Skolaut}, \citenamefont {Yamada},
  \citenamefont {Dup{\'e}},\ and\ \citenamefont
  {Wulfhekel}}]{mougel2020instability}%
  \BibitemOpen
  \bibfield  {author} {\bibinfo {author} {\bibfnamefont {L.}~\bibnamefont
  {Mougel}}, \bibinfo {author} {\bibfnamefont {P.~M.}\ \bibnamefont {Buhl}},
  \bibinfo {author} {\bibfnamefont {R.}~\bibnamefont {Nemoto}}, \bibinfo
  {author} {\bibfnamefont {T.}~\bibnamefont {Balashov}}, \bibinfo {author}
  {\bibfnamefont {M.}~\bibnamefont {Herv{\'e}}}, \bibinfo {author}
  {\bibfnamefont {J.}~\bibnamefont {Skolaut}}, \bibinfo {author} {\bibfnamefont
  {T.~K.}\ \bibnamefont {Yamada}}, \bibinfo {author} {\bibfnamefont
  {B.}~\bibnamefont {Dup{\'e}}}, \ and\ \bibinfo {author} {\bibfnamefont
  {W.}~\bibnamefont {Wulfhekel}},\ }\href@noop {} {\bibfield  {journal}
  {\bibinfo  {journal} {Appl. Phys. Lett.}\ }\textbf {\bibinfo {volume}
  {116}},\ \bibinfo {pages} {262406} (\bibinfo {year} {2020})}\BibitemShut
  {NoStop}%
\bibitem [{\citenamefont {Bogdanov}\ and\ \citenamefont
  {Hubert}(1994)}]{Bogdanov1994a}%
  \BibitemOpen
  \bibfield  {author} {\bibinfo {author} {\bibfnamefont {A.~N.}\ \bibnamefont
  {Bogdanov}}\ and\ \bibinfo {author} {\bibfnamefont {A.}~\bibnamefont
  {Hubert}},\ }\href {\doibase 10.1016/0304-8853(94)90046-9} {\bibfield
  {journal} {\bibinfo  {journal} {J. Magn. Magn. Mater.}\ }\textbf {\bibinfo
  {volume} {138}},\ \bibinfo {pages} {255} (\bibinfo {year}
  {1994})}\BibitemShut {NoStop}%
\bibitem [{\citenamefont {Wilson}\ \emph {et~al.}(2014)\citenamefont {Wilson},
  \citenamefont {Butenko}, \citenamefont {Bogdanov},\ and\ \citenamefont
  {Monchesky}}]{Wilson2014}%
  \BibitemOpen
  \bibfield  {author} {\bibinfo {author} {\bibfnamefont {M.~N.}\ \bibnamefont
  {Wilson}}, \bibinfo {author} {\bibfnamefont {A.~B.}\ \bibnamefont {Butenko}},
  \bibinfo {author} {\bibfnamefont {A.~N.}\ \bibnamefont {Bogdanov}}, \ and\
  \bibinfo {author} {\bibfnamefont {T.~L.}\ \bibnamefont {Monchesky}},\ }\href
  {\doibase 10.1103/PhysRevB.89.094411} {\bibfield  {journal} {\bibinfo
  {journal} {Phys. Rev. B}\ }\textbf {\bibinfo {volume} {89}},\ \bibinfo
  {pages} {094411} (\bibinfo {year} {2014})}\BibitemShut {NoStop}%
\bibitem [{\citenamefont {Horcas}\ \emph {et~al.}(2007)\citenamefont {Horcas},
  \citenamefont {Fern{\'{a}}ndez}, \citenamefont {G{\'{o}}mez-Rodr{\'{i}}guez},
  \citenamefont {Colchero}, \citenamefont {G{\'{o}}mez-Herrero},\ and\
  \citenamefont {Baro}}]{horcas2007}%
  \BibitemOpen
  \bibfield  {author} {\bibinfo {author} {\bibfnamefont {I.}~\bibnamefont
  {Horcas}}, \bibinfo {author} {\bibfnamefont {R.}~\bibnamefont
  {Fern{\'{a}}ndez}}, \bibinfo {author} {\bibfnamefont {J.~M.}\ \bibnamefont
  {G{\'{o}}mez-Rodr{\'{i}}guez}}, \bibinfo {author} {\bibfnamefont
  {J.}~\bibnamefont {Colchero}}, \bibinfo {author} {\bibfnamefont
  {J.}~\bibnamefont {G{\'{o}}mez-Herrero}}, \ and\ \bibinfo {author}
  {\bibfnamefont {A.~M.}\ \bibnamefont {Baro}},\ }\href {\doibase
  10.1063/1.2432410} {\bibfield  {journal} {\bibinfo  {journal} {Rev. Sci.
  Instrum.}\ }\textbf {\bibinfo {volume} {78}},\ \bibinfo {pages} {13705}
  (\bibinfo {year} {2007})}\BibitemShut {NoStop}%
\bibitem [{Vil(2012)}]{Villars2016:sm_isp_sd_1420956}%
  \BibitemOpen
  \href {https://materials.springer.com/isp/crystallographic/docs/sd_1420956}
  {\enquote {\bibinfo {title} {{Fe$_3$GeTe$_2$ Crystal Structure: Datasheet
  from PAULING FILE Multinaries Edition}},}\ } (\bibinfo {year} {2012}),\
  \bibinfo {note} {in SpringerMaterials
  (https://materials.springer.com/isp/crystallographic/docs/sd\_1420956)}\BibitemShut
  {NoStop}%
\bibitem [{\citenamefont {{The JuKKR developers}}(2021)}]{jukkr}%
  \BibitemOpen
  \bibfield  {author} {\bibinfo {author} {\bibnamefont {{The JuKKR
  developers}}},\ }\href {https://jukkr.fz-juelich.de} {\enquote {\bibinfo
  {title} {{The J\"ulich KKR Codes}},}\ } (\bibinfo {year} {2021}),\ \bibinfo
  {note} {https://jukkr.fz-juelich.de}\BibitemShut {NoStop}%
\bibitem [{\citenamefont {Ebert}\ \emph {et~al.}(2011)\citenamefont {Ebert},
  \citenamefont {K{\"o}dderitzsch},\ and\ \citenamefont
  {Min{\'a}r}}]{Ebert2011}%
  \BibitemOpen
  \bibfield  {author} {\bibinfo {author} {\bibfnamefont {H.}~\bibnamefont
  {Ebert}}, \bibinfo {author} {\bibfnamefont {D.}~\bibnamefont
  {K{\"o}dderitzsch}}, \ and\ \bibinfo {author} {\bibfnamefont
  {J.}~\bibnamefont {Min{\'a}r}},\ }\href {\doibase
  10.1088/0034-4885/74/9/096501} {\bibfield  {journal} {\bibinfo  {journal}
  {Rep. Prog. Phys.}\ }\textbf {\bibinfo {volume} {74}},\ \bibinfo {pages}
  {096501} (\bibinfo {year} {2011})}\BibitemShut {NoStop}%
\bibitem [{\citenamefont {Stefanou}\ \emph {et~al.}(1990)\citenamefont
  {Stefanou}, \citenamefont {Akai},\ and\ \citenamefont
  {Zeller}}]{Stefanou1990}%
  \BibitemOpen
  \bibfield  {author} {\bibinfo {author} {\bibfnamefont {N.}~\bibnamefont
  {Stefanou}}, \bibinfo {author} {\bibfnamefont {H.}~\bibnamefont {Akai}}, \
  and\ \bibinfo {author} {\bibfnamefont {R.}~\bibnamefont {Zeller}},\ }\href
  {https://doi.org/10.1016/0010-4655(90)90009-P} {\bibfield  {journal}
  {\bibinfo  {journal} {Comput. Phys. Commun.}\ }\textbf {\bibinfo {volume}
  {60}},\ \bibinfo {pages} {231} (\bibinfo {year} {1990})}\BibitemShut
  {NoStop}%
\bibitem [{\citenamefont {Stefanou}\ and\ \citenamefont
  {Zeller}(1991)}]{Stefanou1991}%
  \BibitemOpen
  \bibfield  {author} {\bibinfo {author} {\bibfnamefont {N.}~\bibnamefont
  {Stefanou}}\ and\ \bibinfo {author} {\bibfnamefont {R.}~\bibnamefont
  {Zeller}},\ }\href {https://doi.org/10.1088/0953-8984/3/39/006} {\bibfield
  {journal} {\bibinfo  {journal} {J. Phys.: Condensed Matter}\ }\textbf
  {\bibinfo {volume} {3}},\ \bibinfo {pages} {7599} (\bibinfo {year}
  {1991})}\BibitemShut {NoStop}%
\bibitem [{\citenamefont {Vosko}\ \emph {et~al.}(1980)\citenamefont {Vosko},
  \citenamefont {Wilk},\ and\ \citenamefont {Nusair}}]{Vosko1980}%
  \BibitemOpen
  \bibfield  {author} {\bibinfo {author} {\bibfnamefont {S.~H.}\ \bibnamefont
  {Vosko}}, \bibinfo {author} {\bibfnamefont {L.}~\bibnamefont {Wilk}}, \ and\
  \bibinfo {author} {\bibfnamefont {M.}~\bibnamefont {Nusair}},\ }\href
  {\doibase 10.1139/p80-159} {\bibfield  {journal} {\bibinfo  {journal} {Can.
  J. Phys.}\ }\textbf {\bibinfo {volume} {58}},\ \bibinfo {pages} {1200}
  (\bibinfo {year} {1980})}\BibitemShut {NoStop}%
\bibitem [{\citenamefont {Perdew}\ \emph {et~al.}(1996)\citenamefont {Perdew},
  \citenamefont {Burke},\ and\ \citenamefont {Ernzerhof}}]{PBE}%
  \BibitemOpen
  \bibfield  {author} {\bibinfo {author} {\bibfnamefont {J.~P.}\ \bibnamefont
  {Perdew}}, \bibinfo {author} {\bibfnamefont {K.}~\bibnamefont {Burke}}, \
  and\ \bibinfo {author} {\bibfnamefont {M.}~\bibnamefont {Ernzerhof}},\ }\href
  {\doibase 10.1103/PhysRevLett.77.3865} {\bibfield  {journal} {\bibinfo
  {journal} {Phys. Rev. Lett.}\ }\textbf {\bibinfo {volume} {77}},\ \bibinfo
  {pages} {3865} (\bibinfo {year} {1996})}\BibitemShut {NoStop}%
\bibitem [{\citenamefont {Liechtenstein}\ \emph {et~al.}(1987)\citenamefont
  {Liechtenstein}, \citenamefont {Katsnelson}, \citenamefont {Antropov},\ and\
  \citenamefont {Gubanov}}]{Liechtenstein1987}%
  \BibitemOpen
  \bibfield  {author} {\bibinfo {author} {\bibfnamefont {A.}~\bibnamefont
  {Liechtenstein}}, \bibinfo {author} {\bibfnamefont {M.}~\bibnamefont
  {Katsnelson}}, \bibinfo {author} {\bibfnamefont {V.}~\bibnamefont
  {Antropov}}, \ and\ \bibinfo {author} {\bibfnamefont {V.}~\bibnamefont
  {Gubanov}},\ }\href {\doibase 10.1016/0304-8853(87)90721-9} {\bibfield
  {journal} {\bibinfo  {journal} {J. Magn. Magn. Mater.}\ }\textbf {\bibinfo
  {volume} {67}},\ \bibinfo {pages} {65} (\bibinfo {year} {1987})}\BibitemShut
  {NoStop}%
\bibitem [{\citenamefont {Udvardi}\ \emph {et~al.}(2003)\citenamefont
  {Udvardi}, \citenamefont {Szunyogh}, \citenamefont {Palot\'as},\ and\
  \citenamefont {Weinberger}}]{PhysRevB.68.104436}%
  \BibitemOpen
  \bibfield  {author} {\bibinfo {author} {\bibfnamefont {L.}~\bibnamefont
  {Udvardi}}, \bibinfo {author} {\bibfnamefont {L.}~\bibnamefont {Szunyogh}},
  \bibinfo {author} {\bibfnamefont {K.}~\bibnamefont {Palot\'as}}, \ and\
  \bibinfo {author} {\bibfnamefont {P.}~\bibnamefont {Weinberger}},\ }\href
  {\doibase 10.1103/PhysRevB.68.104436} {\bibfield  {journal} {\bibinfo
  {journal} {Phys. Rev. B}\ }\textbf {\bibinfo {volume} {68}},\ \bibinfo
  {pages} {104436} (\bibinfo {year} {2003})}\BibitemShut {NoStop}%
\bibitem [{\citenamefont {M\"uller}\ \emph {et~al.}(2019)\citenamefont
  {M\"uller}, \citenamefont {Hoffmann}, \citenamefont {Di\ss{}elkamp},
  \citenamefont {Sch\"urhoff}, \citenamefont {Mavros}, \citenamefont
  {Sallermann}, \citenamefont {Kiselev}, \citenamefont {J\'onsson},\ and\
  \citenamefont {Bl\"ugel}}]{spirit-paper}%
  \BibitemOpen
  \bibfield  {author} {\bibinfo {author} {\bibfnamefont {G.~P.}\ \bibnamefont
  {M\"uller}}, \bibinfo {author} {\bibfnamefont {M.}~\bibnamefont {Hoffmann}},
  \bibinfo {author} {\bibfnamefont {C.}~\bibnamefont {Di\ss{}elkamp}}, \bibinfo
  {author} {\bibfnamefont {D.}~\bibnamefont {Sch\"urhoff}}, \bibinfo {author}
  {\bibfnamefont {S.}~\bibnamefont {Mavros}}, \bibinfo {author} {\bibfnamefont
  {M.}~\bibnamefont {Sallermann}}, \bibinfo {author} {\bibfnamefont {N.~S.}\
  \bibnamefont {Kiselev}}, \bibinfo {author} {\bibfnamefont {H.}~\bibnamefont
  {J\'onsson}}, \ and\ \bibinfo {author} {\bibfnamefont {S.}~\bibnamefont
  {Bl\"ugel}},\ }\href {\doibase 10.1103/PhysRevB.99.224414} {\bibfield
  {journal} {\bibinfo  {journal} {Phys. Rev. B}\ }\textbf {\bibinfo {volume}
  {99}},\ \bibinfo {pages} {224414} (\bibinfo {year} {2019})}\BibitemShut
  {NoStop}%
\bibitem [{\citenamefont {Müller}\ \emph {et~al.}(2021)\citenamefont
  {Müller}, \citenamefont {Sallermann}, \citenamefont {Mavros}, \citenamefont
  {Rhiem}, \citenamefont {Schürhoff}, \citenamefont {Meyer}, \citenamefont
  {Suckert}, \citenamefont {Redies}, \citenamefont {Disselcamp}, \citenamefont
  {Buhl},\ and\ \citenamefont {Blügel}}]{spirit-code}%
  \BibitemOpen
  \bibfield  {author} {\bibinfo {author} {\bibfnamefont {G.~P.}\ \bibnamefont
  {Müller}}, \bibinfo {author} {\bibfnamefont {M.}~\bibnamefont {Sallermann}},
  \bibinfo {author} {\bibfnamefont {S.}~\bibnamefont {Mavros}}, \bibinfo
  {author} {\bibfnamefont {F.}~\bibnamefont {Rhiem}}, \bibinfo {author}
  {\bibfnamefont {D.}~\bibnamefont {Schürhoff}}, \bibinfo {author}
  {\bibfnamefont {I.}~\bibnamefont {Meyer}}, \bibinfo {author} {\bibfnamefont
  {R.}~\bibnamefont {Suckert}}, \bibinfo {author} {\bibfnamefont
  {M.}~\bibnamefont {Redies}}, \bibinfo {author} {\bibfnamefont
  {C.}~\bibnamefont {Disselcamp}}, \bibinfo {author} {\bibfnamefont
  {P.}~\bibnamefont {Buhl}}, \ and\ \bibinfo {author} {\bibfnamefont
  {S.}~\bibnamefont {Blügel}},\ }\href {https://github.com/spirit-code/spirit}
  {\enquote {\bibinfo {title} {Spirit: Spin simulation software},}\ } (\bibinfo
  {year} {2021}),\ \bibinfo {note}
  {\url{https://github.com/spirit-code/spirit}}\BibitemShut {NoStop}%
\bibitem [{\citenamefont {Rüßmann}\ \emph
  {et~al.}(2021{\natexlab{a}})\citenamefont {Rüßmann}, \citenamefont
  {Bertoldo},\ and\ \citenamefont {Blügel}}]{aiida-kkr}%
  \BibitemOpen
  \bibfield  {author} {\bibinfo {author} {\bibfnamefont {P.}~\bibnamefont
  {Rüßmann}}, \bibinfo {author} {\bibfnamefont {F.}~\bibnamefont {Bertoldo}},
  \ and\ \bibinfo {author} {\bibfnamefont {S.}~\bibnamefont {Blügel}},\
  }\href@noop {} {\bibfield  {journal} {\bibinfo  {journal} {Npj Comput.
  Mater.}\ }\textbf {\bibinfo {volume} {7}},\ \bibinfo {pages} {13} (\bibinfo
  {year} {2021}{\natexlab{a}})}\BibitemShut {NoStop}%
\bibitem [{\citenamefont {Rüßmann}\ \emph
  {et~al.}(2021{\natexlab{b}})\citenamefont {Rüßmann}, \citenamefont {{Ribas
  Sobreviela}}, \citenamefont {Sallermann}, \citenamefont {Hoffmann},
  \citenamefont {Rhiem},\ and\ \citenamefont {Blügel}}]{aiida-spirit}%
  \BibitemOpen
  \bibfield  {author} {\bibinfo {author} {\bibfnamefont {P.}~\bibnamefont
  {Rüßmann}}, \bibinfo {author} {\bibfnamefont {J.}~\bibnamefont {{Ribas
  Sobreviela}}}, \bibinfo {author} {\bibfnamefont {M.}~\bibnamefont
  {Sallermann}}, \bibinfo {author} {\bibfnamefont {M.}~\bibnamefont
  {Hoffmann}}, \bibinfo {author} {\bibfnamefont {F.}~\bibnamefont {Rhiem}}, \
  and\ \bibinfo {author} {\bibfnamefont {S.}~\bibnamefont {Blügel}},\
  }\href@noop {} {\bibfield  {journal} {\bibinfo  {journal} {arXiv:2111.15229
  [cond-mat.mtrl-sci]}\ } (\bibinfo {year} {2021}{\natexlab{b}})}\BibitemShut
  {NoStop}%
\bibitem [{\citenamefont {Huber}\ \emph {et~al.}(2020)\citenamefont {Huber},
  \citenamefont {Zoupanos}, \citenamefont {Uhrin}, \citenamefont {Talirz},
  \citenamefont {Kahle}, \citenamefont {Häuselmann}, \citenamefont {Gresch},
  \citenamefont {Müller}, \citenamefont {Yakutovich}, \citenamefont
  {Andersen}, \citenamefont {Ramirez}, \citenamefont {Adorf}, \citenamefont
  {Gargiulo}, \citenamefont {Kumbhar}, \citenamefont {Passaro}, \citenamefont
  {Johnston}, \citenamefont {Merkys}, \citenamefont {Cepellotti}, \citenamefont
  {Mounet}, \citenamefont {Marzari}, \citenamefont {Kozinsky},\ and\
  \citenamefont {Pizzi}}]{aiida}%
  \BibitemOpen
  \bibfield  {author} {\bibinfo {author} {\bibfnamefont {S.~P.}\ \bibnamefont
  {Huber}}, \bibinfo {author} {\bibfnamefont {S.}~\bibnamefont {Zoupanos}},
  \bibinfo {author} {\bibfnamefont {M.}~\bibnamefont {Uhrin}}, \bibinfo
  {author} {\bibfnamefont {L.}~\bibnamefont {Talirz}}, \bibinfo {author}
  {\bibfnamefont {L.}~\bibnamefont {Kahle}}, \bibinfo {author} {\bibfnamefont
  {R.}~\bibnamefont {Häuselmann}}, \bibinfo {author} {\bibfnamefont
  {D.}~\bibnamefont {Gresch}}, \bibinfo {author} {\bibfnamefont
  {T.}~\bibnamefont {Müller}}, \bibinfo {author} {\bibfnamefont {A.~V.}\
  \bibnamefont {Yakutovich}}, \bibinfo {author} {\bibfnamefont {C.~W.}\
  \bibnamefont {Andersen}}, \bibinfo {author} {\bibfnamefont {F.~F.}\
  \bibnamefont {Ramirez}}, \bibinfo {author} {\bibfnamefont {C.~S.}\
  \bibnamefont {Adorf}}, \bibinfo {author} {\bibfnamefont {F.}~\bibnamefont
  {Gargiulo}}, \bibinfo {author} {\bibfnamefont {S.}~\bibnamefont {Kumbhar}},
  \bibinfo {author} {\bibfnamefont {E.}~\bibnamefont {Passaro}}, \bibinfo
  {author} {\bibfnamefont {C.}~\bibnamefont {Johnston}}, \bibinfo {author}
  {\bibfnamefont {A.}~\bibnamefont {Merkys}}, \bibinfo {author} {\bibfnamefont
  {A.}~\bibnamefont {Cepellotti}}, \bibinfo {author} {\bibfnamefont
  {N.}~\bibnamefont {Mounet}}, \bibinfo {author} {\bibfnamefont
  {N.}~\bibnamefont {Marzari}}, \bibinfo {author} {\bibfnamefont
  {B.}~\bibnamefont {Kozinsky}}, \ and\ \bibinfo {author} {\bibfnamefont
  {G.}~\bibnamefont {Pizzi}},\ }\href@noop {} {\bibfield  {journal} {\bibinfo
  {journal} {Sci. Data}\ }\textbf {\bibinfo {volume} {7}},\ \bibinfo {pages}
  {300} (\bibinfo {year} {2020})}\BibitemShut {NoStop}%
\end{thebibliography}


\newpage

\end{document}